\documentclass[useAMS,usenatbib,aas_macros]{mn2e}
\usepackage{aas_macros}
 \usepackage{epsfig}
 \usepackage{aastex_hack}
 \usepackage{multirow}

\def\eq#1{Eq.~(\ref{#1})}
\def\Eq#1{Eq.~(\ref{#1})}
\def\sec#1{\S\ref{#1}}

\def\Fig#1{Fig.~\ref{#1}}
\def\fig#1{Fig.~\ref{#1}}

\def\Tab#1{Table \ref{#1}}
\def\tab#1{Table \ref{#1}}
\def\twotabs#1#2{Tables \ref{#1} and \ref{#2}}

\def\tfnm#1{$^{\it #1}$}
\def\tfnt#1#2{{\scriptsize $^{\it #1}$#2}}

\def\tref#1{{\scriptsize {\bf References.} --- #1}}

\def\colh#1{\multicolumn{1}{c}{#1}}
\def\mcol#1#2#3{\multicolumn{#1}{#2}{#3}}
\def\mrow#1#2#3{\multirow{#1}{#2}{#3}}

\def\ie{i.e. }
\def\eg{e.g.}
\def\be{\begin{equation}}
\def\ee{\end{equation}}
\def\prop{\propto}
\def\ifm#1{\relax\ifmmode#1\else$\mathsurround=0pt #1$\fi}
\def\kms{\ifmmode\,{\rm km}\,{\rm s}^{-1}\else km$\,$s$^{-1}$\fi}

\def\kpc{{\rm kpc}}

\def\msun{{\rm  M}_{\odot} }
\def\Msun{{\rm  M}_{\odot} }

\def\ltsima{$\; \buildrel < \over \sim \;$}

\def\lsim{\lower.5ex\hbox{\ltsima}}
\def\gtsima{$\; \buildrel > \over \sim \;$}
\def\gsim{\lower.5ex\hbox{\gtsima}}

\def\Ms{M_{\rm *}}

\def\vrot{V_{\rm rot}}
\def\Vrot{V_{\rm rot}}

\def\MtLv{M_{\rm tot}/L_{V}}
\def\MSL{M_{*}/L}
\def\Rs{R_{*}}

\def\sfr{\dot{M}_*}
\def\mus{\mu_*}
\def\Mbar{M_{\rm bar}}
\def\HI{\ion{H}{1} }
\def\HII{\ion{H}{2} }
\def\MHI{M_{\rm HI}}

\def\logMstarv_b{\log \Ms}
\def\mustarv_b{\log \mus}
\def\logv4{\log V}
\def\logZ{\log Z}
\def\logrp{\log \Rs}
\def\logsfr{\log \sfr}

\def\sbis{s_{\rm bis}}
\def\bbis{b_{\rm bis}}
\def\sigbis{\sigma_{\rm bis}}

\def\s1d{s_{\rm 1D}}
\def\b1d{b_{\rm 1D}}
\def\sig1d{\sigma_{\rm 1D}}

\def\MLBCvi{0.7}
\def\MLBCvs{1.6}
\def\MLBCvt{1.0}
\def\MLBJvi{0.8}
\def\MLBJvs{1.3} 
\def\MLBJvt{1.0} 
\def\MLBCbi{0.5}
\def\MLBCbs{1.5}
\def\MLBCbt{0.7}
\def\MLBJbi{0.6} 
\def\MLBJbs{1.1} 
\def\MLBJbt{0.8} 

\def\elogMs{0.17 }
\def\erangeMs{4}
\def\elogmus{0.17 }
\def\erangemus{7}

\def\erangeV{4}
\def\elogZ{0.2 }
\def\erangeZ{10}

\def\rmus{0.81}
\def\decMs{4.5 } 
\def\decmus{3 }
\def\scmus{0.61 \pm 0.05}
\def\scmusde{0.74 \pm 0.05}

\def\musoff{0.5 } 

\def\scrp{0.33 \pm 0.02}
\def\scrpde{0.26 \pm 0.05}
\def\scrpdi{0.41 \pm 0.04}
\def\rMsrp{0.85}

\def\scZ{0.40 \pm 0.04}

\def\scOx{0.39 \pm 0.04}
\def\rMZ{0.76}
\def\rMZs{0.93}
\def\rMZi{0.88}

\def\scV{0.27 \pm 0.01}

\def\scVdi{0.26 \pm 0.02}
\def\rV{0.94}
\def\scVdirev{3.8 \pm 0.3}
\def\BTFdIrev{3.3 \pm 0.2}

\def\scsfr{1.18 \pm 0.08}
\def\rsfr{0.96}

\title[The Dwarf Galaxy Fundamental Line]{Scaling Relations and the Fundamental Line of the Local
  Group Dwarf Galaxies}

\author[J. Woo, S. Courteau \& A. Dekel]
  {Joanna Woo$^1$, St\'ephane Courteau$^2$ \& Avishai Dekel$^1$\\
\\
$^1$Racah Institute of Physics, The Hebrew University, Jerusalem 91904 Israel\\
$^2$Department of Physics and Astronomy, Queen's University, Kingston, Ontario, Canada\\
joaw@phys.huji.ac.il; courteau@astro.queensu.ca; dekel@phys.huji.ac.il}

\begin{document}

\pagerange{\pageref{firstpage}--\pageref{lastpage}} \pubyear{2006}

\maketitle

\label{firstpage}

\begin{abstract}

 We study the scaling relations between global properties of  
 dwarf galaxies in the Local Group.  In addition to quantifying  
 the correlations between pairs of variables, we explore the  
 ``shape'' of the distribution of galaxies in log parameter  
 space using standardised Principal Component Analysis (PCA),  
 the analysis is performed first in the 3-D structural parameter  
 space of stellar mass $\Ms$, internal velocity $V$ and characteristic 
 radius $\Rs$ (or surface brightness $\mus$).  It is then extended  
 to a 4-D space that includes a stellar-population parameter such 
 as metallicity $Z$ or star formation rate $\sfr$.  We find that  
 the Local-Group dwarfs basically define a one-parameter ``Fundamental  
 Line''(FL), primarily driven by stellar mass, $\Ms$.  A more detailed  
 inspection reveals differences between the star-formation properties  
 of dwarf irregulars (dI's) and dwarf ellipticals (dE's), beyond the  
 tendency of the latter to be more massive.  In particular, the  
 metallicities of dI's are typically lower by a factor of 3 at a  
 given $\Ms$ and they grow faster with increasing $\Ms$, showing  
 a tighter FL in the 4-D space for the dE's.  The structural  
 scaling relations of dI's resemble those of the more massive  
 spirals, but the dI's have lower star-formation rates for a  
 given $\Ms$ which also grow faster with increasing $\Ms$.  On  
 the other hand, the FL of the dE's departs from the fundamental  
 plane of bigger ellipticals.  While the one-parameter nature of  
 the FL and the associated slopes of the scaling relations are  
 consistent with the general predictions of supernova feedback  
 \citep{dek03}, the differences between the FL's of the  
 dE's and the dI's remain a challenge and should serve as a  
 guide for the secondary physical processes responsible for  
 these two types. 
 
\end{abstract}


\section{Introduction}
\label{intro}

\subsection{Scaling Relations, Fundamental Distributions
  and Galaxy Formation in Giant Galaxies}

In the standard $\Lambda$CDM picture for galaxy formation, galaxies
form from the gas that cools within a dark matter halo into a disc of
stars whose
specific angular momentum is conserved \citep{fal80}.  Since the gas
initially shares the spatial distribution and dynamical properties of
the dark matter halo, the structural properties of the resulting
gaseous disc, specifically the mass of the disc of stars $\Ms$, the size
of the disc $\Rs$ and its rotation velocity $\Vrot$, are expected to
correlate strongly with those of the halo.  The halo itself is expected to be
characterised by the virial theorem, \ie 
$M_{\rm vir} \prop V_{\rm vir}^3 \prop R_{\rm vir}^3$.  The actual
scalings of the stellar disc will be affected by the physics of the
baryon gas and its interactions with the dark matter halo
\citep{dut07}.  Thus the observed scaling relations between these
structural  
parameters in galaxies help to shed light on the physical processes 
governing their formation.

Several recent studies of scaling relations of galaxies have contributed 
to significant improvements in our appreciation of galaxy formation physics.
For example, the scaling relation between $\Ms$ and the stellar surface 
density $\mus$ (related to $\Rs$) for 123,000 galaxies of all types in 
the Sloan Digital Sky Survey (SDSS) shows a transition at 
$\sim3 \times 10^{10 } \Msun$ \citep{kau03a}.  
Above the transition, $\mus$ only weakly depends on $\Ms$ while 
below it, $\mus$ scales as $\Ms^{0.6}$, consistent with
simple theoretical predictions of supernova feedback (SNF)
\citep{dek86,dek03}.  The transition itself can also be explained
by considering the energy constraints of gas ejection by
supernovae \citep{dek86}.

The interpretation of the scaling relations between $\Vrot$ and 
luminosity $L$, and $\Rs$ and $L$ for $\sim1300$ local 
spiral galaxies by \cite{cou07} also suggests that 
the interaction of the disc and halo during galaxy formation
is a fundamental constraint to $\Lambda$CDM models
\citep{dut07}. 

The study of scaling relations for elliptical galaxies has revealed the 
existence of a "fundamental plane'' (FP) \citep{djo87,dre87,ber03}.  
These galaxies lie on a plane in the parameter space spanned by 
the logarithms of $L$, $R$, and the velocity dispersion $\sigma$.  
The projections of this plane onto the $\sigma-L$ and $R-L$ planes 
are known as the Faber-Jackson \citep{fab76} and the Kormendy
\citep{kor77} relations.  Recently, \cite{zar06,zar06b,zar07} proposed
that the slight tilt of the FP can be straightened out essentially by adding
mass-to-light ratio as an extra dimension to the parameter space.
\cite{dek06} used hydrodynamical simulations 
to show how the properties of the fundamental plane of elliptical
galaxies, including its tilt with respect to virial theorem predictions, 
can be explained by considering the dissipative properties 
of the major mergers that form them.

Besides the structural parameters ($\Ms$, $\mus$, $\Rs$, $V$), 
the physics of the gaseous systems that later become galaxies
will also affect the global metallicity $Z$, star formation history, 
and consequently the current star formation rate $\sfr$ (SFR) of 
these galaxies.  Since $Z$ and $\sfr$ are related to star formation 
we call these the ``SF'' quantities for brevity.  

Studies that involve the SF parameters with stellar mass have 
also yielded insight into the gas and stellar population physics of 
galaxies.  For example, \citet{tre04} show from SDSS measurements 
of galaxies with $10^{8.5} \Msun < \Ms < 10^{10.5} \Msun$ that $Z$ 
decreases systematically with decreasing $\Ms$, while higher mass 
systems only show a weak dependence on $\Ms$.   These scalings are 
consistent with SNF model predictions \citep{dek03,tas03}.
As another example, \cite{bri04} show trends of current star formation
rate $\sfr$ with $\Ms$, and \cite{kau04} show environmental dependence 
of this relation.  This poorly explored scaling relation may reveal insight 
into the effect of the disc self-gravity on gas processes that combine to
form stars.

\subsection{Dwarf Galaxies}

These investigations of the scaling relations and the FP have led to
major advancements in our understanding of galaxy formation with
respect to giant galaxies, and this motivates a similar investigation
of dwarf galaxies.  Dwarf galaxies are the central players in one of
the most puzzling problems in the $\Lambda$CDM picture of galaxy
formation, namely, the ``missing dwarf problem.''  This relates to the
apparent discrepancy between the predicted distribution of halo masses
at the low-mass end and the relatively few dwarf galaxies actually
observed - see for example (\citealp{kly99,moo99,sto02}).  Similarly,
the luminosity function of galaxies below the Schechter's
characteristic $L_*$ is observed to be flatter than predicted in
$\Lambda$CDM cosmology, implying that the ratio of stellar mass to
total mass decreases with decreasing $M_{\rm tot}$.  \citet{dek03}
used this systematic variation in $\Ms/M_{\rm tot}$ to predict the
scaling relations of $\mus \prop \Ms/R^2$ and $V$ with $\Ms$ for dwarf
galaxies in the SNF scenario.  (See also \citealp{mal02} who show how
SNF helps solve the angular momentum problem, and \citealp{dekbir06} who
detail the role of SNF in small galaxies).  A dependence of the ratio
of $\Ms/M$ on $M$ for low-mass galaxies, and thus a dependence of the
overall star formation efficiency on $M$, should also have
consequences on the scaling relations of the SF quantities with $\Ms$.

The theoretical developments concerning the scaling relations between
the properties of dwarf galaxies motivate an investigation of the
fundamental distribution of the dwarf galaxies in parameter space. 
For example, \cite{ben92} plot the galaxies in what they call
``$\kappa$''-space. 
\citet{pra02} (hereafter PB02) find a  ``fundamental line'' (FL) 
of dwarf galaxies in the space of $M_{\rm tot}/L$, metallicity [Fe/H], and 
the central surface brightness $\Sigma_V$, which they explain with a 
simple chemical enrichment model.  PB02 find that this FL is independent 
of Hubble type \citep[see also][]{sim06}.   Similarly, \cite{vad05b} find a 
fundamental plane of dI's in the parameter space of absolute $K$ magnitude
$M_K$, central surface brightness $\Sigma_K$ and \HI line width $W$.
The same authors find that blue compact dwarfs (BCD's) 
lie on the same fundamental plane as dI's \citep{vad06}.  
\cite{car92} and \cite{zar06b} find that the dE's lie in the same ``fundamental
manifold'', the latter putting them on the same manifold as larger
elliptical galaxies.  However these fundamental 
distributions lie in parameter spaces whose axes are various combinations
of structural and SF quantities.  In order to better understand the
underlying physics behind these fundamental distributions, a
disentangling of these axes into more ``fundamental'', or linearly
independent quantities is in order. 

The scaling relations between the structural and SF quantities
in the dwarf galaxies were first presented as luminosity-dependent relations; 
\eg $V$-$L$: \citealp{fab76} and \citealp{tul77}; $\Sigma-L$: 
\citealp{ben92}; [Fe/H]-$L$, [O/H]-$L$: \citealp{zar94} and
\citealp{ric98}. However, for the purpose of constraining and comparing 
with physical models, the use of physical quantities is preferable. 

We wish to improve on these investigations.  Our goals are two-fold:
(i) present the scaling relations of dwarf galaxies in physical quantities;
and (ii), find a fundamental distribution in a parameter space spanned
by linearly independent physical quantities that can constrain physical models.
We will show that in such a parameter space the dwarf galaxies lie in
a linear distribution.  Such a distribution, 
or FL, would be described by a one-parameter family of equations 
which narrowly constrains the possible physical scenarios of their 
origin.  We will show this parameter space to include both
structural and SF quantities.

The dwarf galaxies of the Local Group (LG), whose stellar masses range from
as low as $10^{5.5} \Msun$ to the SDSS transition scale of $10^{10.5} \Msun$, 
provide an excellent opportunity to achieve our goals.  
In this paper, we compile data for LG dwarf galaxies and present scaling 
relations for the structural and SF properties of the dwarf galaxies
in physical units.  We use a Principal Component Analysis (PCA) to 
search for  linear distributions in structural and SF space and quantify their
tightness.  We identify the space, involving both structural and SF quantities, 
in which the LG dwarf galaxies form a fundamental line.

The outline of this paper is as follows:
the data base is described in \sec{data}, followed in \sec{stellarmass} by  
our derivation of stellar mass and other global properties.
In \sec{tools} we describe the tools of our analysis, namely the 
     fitting prescription and principal component analysis.
In \sec{stranalysis}, we present the scaling relations 
     of LG dwarf galaxies in structural parameter space.
In \sec{speanalysis}, we augment the scaling relations with SF
     parameters and mass to derive the fundamental linear distributions 
     of the dwarf galaxies in the parameters space of structural + SF parameters.
Our results are discussed and summarised in \sec{conc}.

\section{Data}
\label{data}


 \begin{table*}
 \begin{minipage}{175mm}
 \caption{\small Data for the Local Group Dwarf Galaxies \label{table}}
 \begin{scriptsize}
 \begin{tabular}{lccccccccccr}
 \hline\hline
 \colh{Name}     & \colh{Type\tfnm{a}} & \colh{$\MSL_V$\tfnm{b}} & \colh{$\log \Ms$\tfnm{c}}
 & \colh{$\log \mus$\tfnm{d}}       & \colh{$\log \Rs$\tfnm{e}}
 & \colh{$V$\tfnm{f}} & \colh{$\MHI$\tfnm{g}} & \colh{$\log Z$\tfnm{h}}
 & \colh{SFR\tfnm{i}} & \colh{($B-V$)$_o$\tfnm{j}} & \colh{Ref.} \\
 \hline
M 33            & dI &   1.2   &   9.55   &    8.51   &    0.37 $\pm$   0.01  &     135   $\pm$ 15      &    1800       &   -2.7   &   0.24       &   0.61       &   2,4,6,10,13,\\
                &    &         &          &           &                       &                          &               &          &              &              &   16,18,19,20   \\
LMC             & dI &   0.7   &   9.19   &    8.14   &    0.17 $\pm$   0.02  &       72   $\pm$  7      &     500       &   -2.3   &   0.26       &   0.45       &   1,7,11,16,19              \\
NGC 55          & dI &   0.8   &   9.04   &    8.15   &    0.17 $\pm$   0.05  &       86   $\pm$  3      &    1404       &     -    &   0.18       &   0.49       &   14,16                     \\
SMC             & dI &   0.8   &   8.67   &    7.63   &    0.15               &       60                 &     420       &   -2.9   &   0.046      &   0.48       &   7,11,16,17,19             \\
M 32            & dE &   1.3   &   8.66   &   12.09   &            -          &       96   $\pm$ 19      &       2.5     &   -2.8   &       -      &   0.93       &   7,14,16                   \\
NGC 205         & dE &   1.4   &   8.65   &    8.54   &   -0.39               &       89   $\pm$ 15      &       0.39    &   -2.2   &       -      &   0.64       &   7,14,16,19                \\
IC 10           & dI &   0.9   &   8.30   &    8.37   &   -0.44 $\pm$   0.08  &       47   $\pm$  4      &      98       &   -3.0   &   0.06       &   0.53       &   7,14,15,19                \\
NGC 6822        & dI &   0.8   &   8.23   &    7.91   &   -0.46 $\pm$   0.07  &       51   $\pm$  3      &     140       &   -2.9   &   0.021      &   0.55       &   7,14,16,19                \\
NGC 147         & dE &   1.6   &   8.17   &    8.12   &   -0.34 $\pm$   0.07  &       44   $\pm$ 10      &       0.005   &   -2.8   &       -      &   0.75       &   7,14,16,19                \\
NGC 185         & dE &   1.0   &   8.14   &    8.51   &   -0.51 $\pm$   0.08  &       46   $\pm$  4      &       0.13    &   -2.5   &       -      &   0.74       &   7,14,16,19                \\
NGC 3109        & dI &   0.8   &   8.13   &    6.84   &    0.09               &       68   $\pm$  4      &     820       &   -3.4   &   0.02       &   0.45       &   7,14,16                   \\
IC 1613         & dI &   1.0   &   8.03   &    7.42   &    0.05 $\pm$   0.24  &       37   $\pm$  3      &      58       &   -3.1   &   0.003      &   0.69       &   7,14,16,19                \\
WLM             & dI &   0.9   &   7.65   &    8.36   &   -0.04 $\pm$   0.03  &       23   $\pm$  2      &      63       &   -3.1   &   0.0011     &   0.58       &   7,14,16,19                \\
IC 5152         & dI &   0.5   &   7.54   &     -     &            -          &       38   $\pm$  4      &      67       &   -3.1   &       -      &   0.32       &   7,14,16                   \\
Sextans B       & dI &   0.8   &   7.54   &     -     &            -          &       38   $\pm$  9      &      44       &   -3.8   &   0.002      &   0.49       &   7,14,16,19                \\
Sextans A       & dI &   0.5   &   7.43   &    6.82   &   -0.20 $\pm$   0.09  &       33   $\pm$  2      &      54       &   -3.6   &   0.002      &   0.34       &   7,14,16,19                \\
Fornax          & dE &   1.2   &   7.27   &    7.29   &   -0.39               &       20   $\pm$  3      &       0.7     &   -2.9   &       -      &   0.61       &   7,14,16,19                \\
EGB 0427+63     & dI &   1.0   &   6.96   &    6.99   &            -          &       33   $\pm$ 10      &      17       &     -    &   0.0004     &   0.55       &   14,16                     \\
And I           & dE &   1.6   &   6.85   &    7.01   &   -0.47 $\pm$   0.02  &          -               &       0.096   &   -3.1   &       -      &   0.75       &   7,14,16,19                \\
Pegasus         & Tr &   1.0   &   6.83   &     -     &            -          &       17   $\pm$  3      &       3.4     &   -3.7   &   0.0003     &   0.55       &   7,14,16,19                \\
And VII         & dE &   0.9   &   6.67   &    7.10   &   -0.54               &          -               &       -       &   -3.2   &       -      &   0.51       &   3,7,8,16                  \\
Leo I           & dE &   0.9   &   6.66   &    7.57   &   -0.85 $\pm$   0.05  &       17   $\pm$  3      &       0.009   &   -3.1   &       -      &   0.76       &   7,14,16,19                \\
And II          & dE &   1.0   &   6.63   &    6.75   &   -0.51 $\pm$   0.02  &       18   $\pm$  5      &       -       &   -3.2   &       -      &   0.54       &   5,7,14,16                 \\
UKS 2323-326    & dI &   0.6   &   6.51   &    6.50   &            -          &          -               &       -       &     -    &       -      &   0.39       &   14,16                     \\
GR 8            & dI &   0.7   &   6.42   &    7.49   &   -0.81 $\pm$   0.07  &       21   $\pm$  6      &       9.6     &   -3.7   &   0.0007     &   0.34       &   7,14,16                   \\
Antlia          & Tr &   1.0   &   6.41   &    6.84   &   -0.37               &       12   $\pm$  3      &       0.72    &   -3.6   &   0.00028    &   0.52       &   7,14,16,19                \\
Sag DIG         & dI &   0.4   &   6.36   &    6.44   &            -          &       14   $\pm$  4      &       8.6     &   -4.0   &   0.000067   &   0.28       &   7,14,16                   \\
And III         & dE &   1.8   &   6.27   &    6.70   &   -0.78 $\pm$   0.04  &          -               &       0.09    &   -3.4   &       -      &   0.54       &   7,14,16                   \\
Leo A           & dI &   0.5   &   6.26   &     -     &            -          &       18                 &       7.6     &   -3.8   &   0.000032   &   0.31       &   7,14,16,19                \\
DDO 210         & Tr &   0.9   &   6.25   &    7.32   &            -          &       13   $\pm$  3      &       2.7     &   -3.6   &       -      &   0.25       &   7,12,14,16                \\
And VI          & dE &   0.5   &   6.17   &    6.55   &   -0.53 $\pm$   0.03  &          -               &       -       &   -3.4   &       -      &   0.34       &   3,7,8,9,16                \\
Leo II          & dE &   1.6   &   6.16   &    7.16   &   -1.05               &       13   $\pm$  2      &       0.03    &   -3.3   &       -      &   0.63       &   7,14,16,19                \\
Phoenix         & Tr &   1.8   &   6.10   &     -     &            -          &       17   $\pm$  3      &       0.17    &   -3.6   &       -      &   0.59       &   7,14,16                   \\
Sculptor        & dE &   1.7   &   6.08   &    7.31   &   -0.77 $\pm$   0.02  &       12   $\pm$  2      &       0.09    &   -3.2   &       -      &   0.68       &   7,14,16,19                \\
Tucana          & dE &   1.6   &   5.97   &    6.76   &   -0.91 $\pm$   0.08  &          -               &       0.015   &   -3.4   &       -      &   0.67       &   7,14,16,19                \\
Draco           & dE &   1.8   &   5.96   &    7.07   &   -0.99               &       16   $\pm$  3      &       0.008   &   -3.7   &       -      &   0.92       &   7,14,16,19                \\
Sextans         & dE &   1.6   &   5.93   &    6.28   &   -0.51 $\pm$   0.11  &       12   $\pm$  2      &       0.03    &   -3.6   &       -      &   0.65       &   7,14,16,19                \\
LGS 3           & Tr &   1.0   &   5.86   &    6.68   &   -0.85               &       17.5 $\pm$  0.6    &       0.2     &   -3.4   &       -      &   0.70       &   7,14,16,19                \\
UMi             & dE &   1.9   &   5.75   &    6.63   &   -0.79 $\pm$   0.14  &       20   $\pm$  2      &       0.007   &   -3.6   &       -      &   1.27       &   7,14,16,19                \\
Carina          & dE &   1.0   &   5.71   &    6.38   &   -0.82               &       13   $\pm$  3      &       0.001   &   -3.5   &       -      &   0.64       &   7,14,16,19                \\
And V           & dE &   1.1   &   5.60   &    6.67   &   -0.84               &          -               &       -       &   -3.6   &       -      &   0.57       &   3,7,16                    \\
 \hline
 \end{tabular}
 \end{scriptsize}
 \tfnt{a}{Morphological type simplified from \citet{mat98},
 \citet{van00}, and \citet{gre03} as described in the text. \\}
 \tfnt{b}{Our derived values of $\MSL_V$ for the ``b'' set as described
   in the text. \\}
 \tfnt{c}{Stellar mass, in units of $\Msun$, derived as described
 in text. The typical uncertainty: \elogMs dex. \\}
 \tfnt{d}{Central stellar surface mass density, in units of $\Msun$ kpc$^{-2}$,
 derived as described in text. The typical uncertainty: \elogmus dex. \\}
 \tfnt{e}{Exponential radius in \kpc. \\}
 \tfnt{f}{Estimated velocity, in \kms, defined as the maximum
 of $v_{\rm rot}$ and $2\sigma$. \\}
 \tfnt{g}{\HI gas mass in units of $10^6 \Msun$
 taken from \citet{gre03}.\\}
 \tfnt{h}{Following \sec{data}, log($Z$/0.019) = [Fe/H]. Typical uncertainty:
 \elogZ dex (Eva Grebel, private communication). \\}
 \tfnt{i}{Current star formation rate in $\Msun$ year$^{-1}$. \\}
 \tfnt{j}{De-reddened $B-V$ colour. Reddening values are from \citet{sch98}
 except for IC 10 whose source is \citet{ric01}. \\}
 \tref{
 (1) \citet{alv00}; 
 (2) \citet{bag98}; 
 (3) \citet{cal99}; 
 (4) \citet{cor00}; 
 (5) \citet{cot99}; 
 (6) \citet{eng03}; 
 (7) \citet{gre03}; 
 (8) \citet{gre99}; 
 (9) \citet{hop99}; 
 (10) \citet{jac70}; 
 (11) \citet{lar00}; 
 (12) \citet{lee99}; 
 (13) \citet{lee02}; 
 (14) \citet{mat98}; 
 (15) \citet{ric01}; 
 (16) \citet{sch98}; 
 (17) \citet{sta04}; 
 (18) \citet{tie04}; 
 (19) \citet{van00}; 
 (20) \citet{zar89}; 
 }
 \end{minipage}
 \end{table*}


Our compilation of relevant global parameters for LG dwarf galaxies
is presented in \Tab{table}.  
It includes the:
\begin{enumerate}
\renewcommand{\labelenumi}{(\alph{enumi})}
\item
stellar mass-to-light-ratio $\MSL$;
\item
stellar mass $\Ms$ in solar masses;
\item
central surface density $\mus$ in $\Msun ~ \kpc^{-2}$;
\item
exponential scale length $\Rs$ in $\kpc$;
\item
characteristic rotational velocity $V$ in \kms, taken 

\hspace{13pt} to represent the halo potential well;
\item
\HI gas mass in units of $10^6 \Msun$;
\item
mean metallicity $Z$;
\item
current star formation rate $\sfr$;
\item
colour $B-V$.
\end{enumerate}

These global parameters
are derived primarily from the compilations of \citet[hereafter M98]{mat98}, 
\citet[hereafter vdB00]{van00}, and \citet[hereafter GGH03]{gre03}. 
GGH03 also provide absolute $V$ magnitudes, surface brightnesses and 
metallicities ([Fe/H]).    
Of all the galaxies listed in these sources, 
we excluded the Sagittarius dwarf spheroidal given its strong tidal 
perturbations with the Milky Way.


We have classified the LG dwarfs into three basic types, simplifying the 
classification system adopted by M98, vdB00, and GGH03.
These are dwarf irregulars (dI), comprising both dIrr and Irr galaxies, 
but mostly the former; 
``early-type'' dwarf galaxies (dE), comprising both dwarf ellipticals and 
dwarf spheroidals, but mostly the latter;
and transition galaxies (Tr), normally classified as ``dIrr/dSph''.  
The dE galaxies tend to be less massive, 
less gas-rich, and more metal-rich than the dI's.  
The Tr galaxies appear to be a morphological 
transition between dIrr's and dSph's.  The Tr's tend to be currently
forming stars like the dIrr's, but are more similar to dSph's in size
and shape \citep[\eg][]{san91}.  

The first four parameters ($\Ms$, $\mus$, $\Rs$, $V$) in the above 
list are the structural parameters, while $Z$ and $\sfr$ are SF 
parameters since they are related to the star formation 
of the galaxies.  We describe our derivation of these parameters below.

\subsection{Stellar Mass Derivation}
\label{stellarmass}

\subsubsection{Methods}

We consider two basic methods to estimate the stellar
mass-to-light ($\MSL$) ratio of a galaxy based either on colours or
inferred SFH's.
We use a combination of both methods in two different ways to 
produce two data sets.  


The first method uses the calibration of colours with model $\MSL$ by
\citet{bel01} and \citet{bel03}.  Hereafter,
we will refer to this colour-$\MSL$ calibration as ``BdJ''. 
BdJ generated model galaxies with model star formation histories (SFH)
following a variety of star formation laws based on the 
simple stellar population (SSP) models of 
\citet[hereafter BC03]{bru03}. 


Our second method 
uses inferred SFH's rather than colours to calculate the $\MSL$ 
ratio directly from the SSP models from BC03. 
The SSP models describe the photometric evolution of single, instantaneous
bursts of star formation with different metallicities.  M98 provides SFH's 
in the form of histograms of relative star formation rate (SFR) as a function 
of age for many of the dwarf galaxies (see his Fig. 8).  Sampling and 
systematic errors in the SFH's are not easily estimated.  However, M98 
noted that for the majority of the galaxies, the duration and relative 
strength of most of the star formation (SF) episodes are ``fairly well
determined'',  
while only a few galaxies (GR 8, Tucana, DDO 210, Sextans B, and M 32)
are dominated by SF episodes whose duration or relative strength are
``greatly uncertain'' (see the caption of his Fig. 8).

To calculate $\MSL$ using SFH's from M98, metallicities and the 
SSP we follow the {\it isochrone synthesis} technique of \cite{bru93} 
which assumes that a stellar population with a given SFH can be decomposed
into a series of instantaneous bursts of SF.  The details of our calculation
are presented in Appendix A.

\subsubsection{Comparison of Methods}
\label{compare}

\begin{figure*}
\epsscale{2.2}
\plottwo{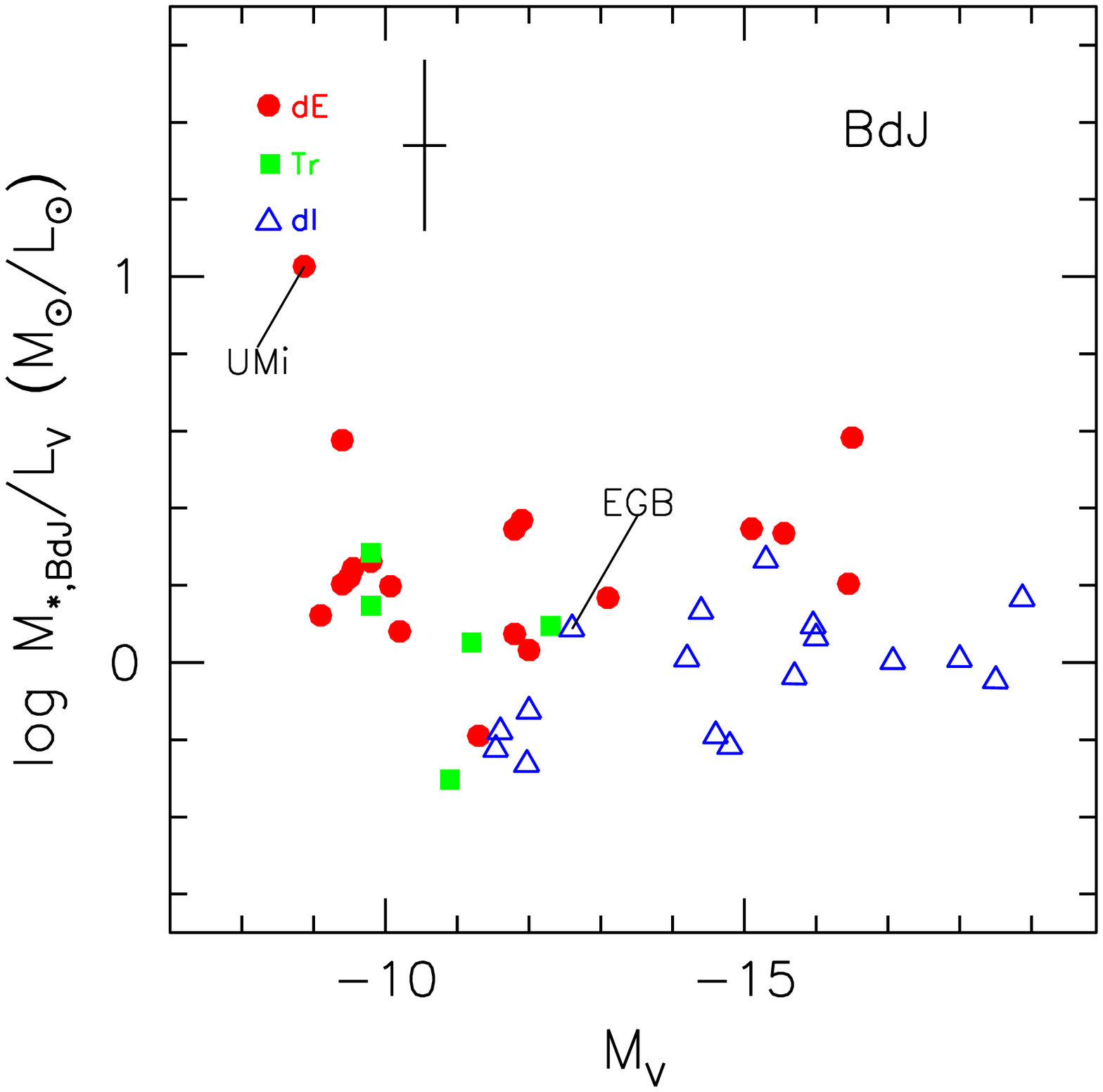}{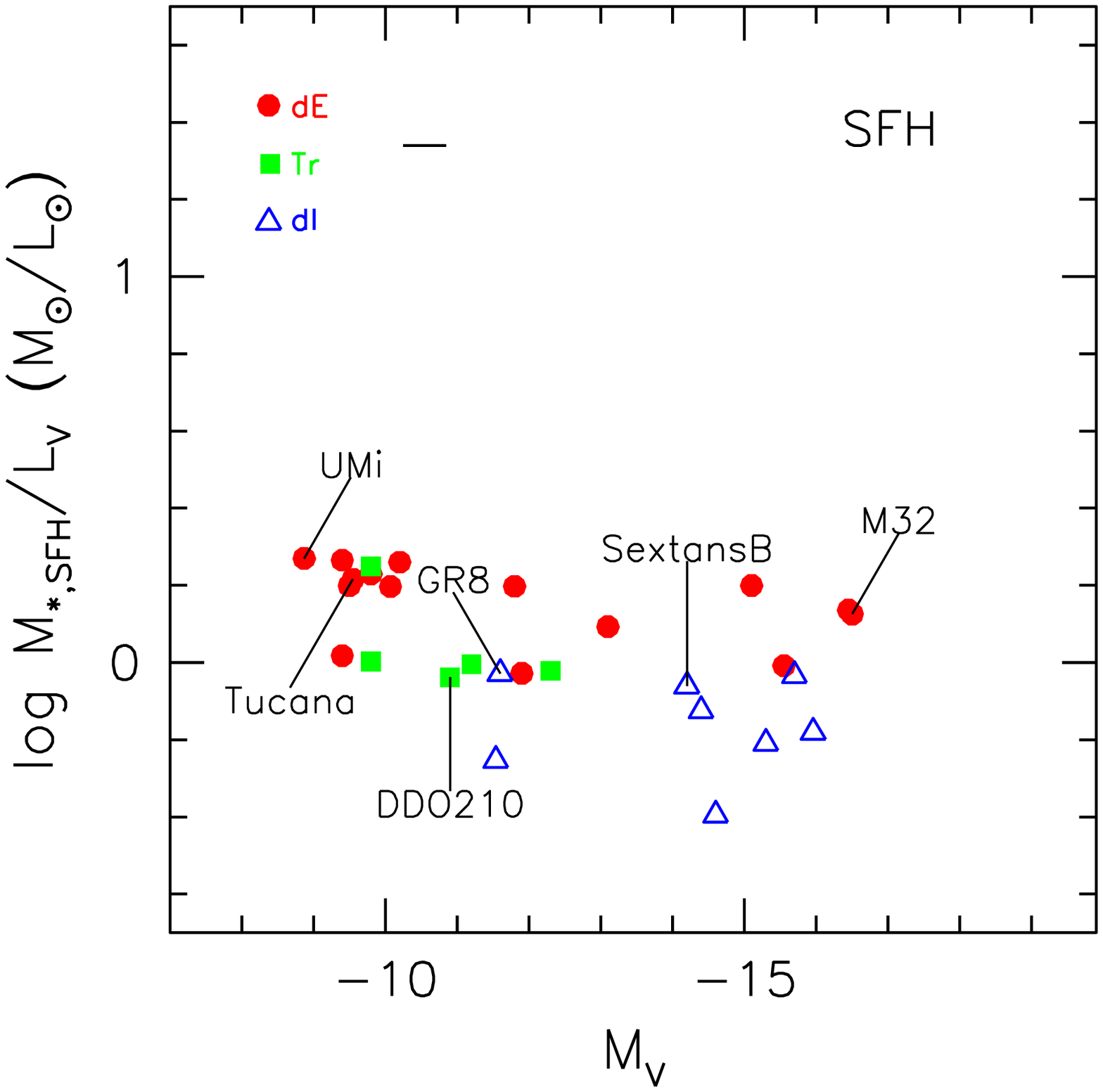}
\caption{The BdJ and SFH stellar mass-to-light ratios, in the
$V$-band, as a function of $M_V$.}
\label{MstoL}
\end{figure*}

The $\MSL_V$ ratios calculated with the two methods above 
are plotted against absolute $V$ magnitude in \Fig{MstoL}. 
There is no apparent trend in either plot.
The $\MSL$ values calculated through BdJ's colour-$\MSL$ relation
are more scattered than those calculated from inferred SFH's, likely
because BdJ's colour-$\MSL$ relations were derived for gaseous discs,
and are not ideal for early-type dwarf galaxies.

Immediately apparent in both plots is an offset in $\MSL$
between dI and dE galaxies.  The dE's have a higher $\MSL$ on average than 
dI's, as expected from their older, redder, stellar populations.  
Median values for $\MSL$ according to type are given in 
\tab{avgML}.

\begin{figure}
\epsscale{1.0}
\plotone{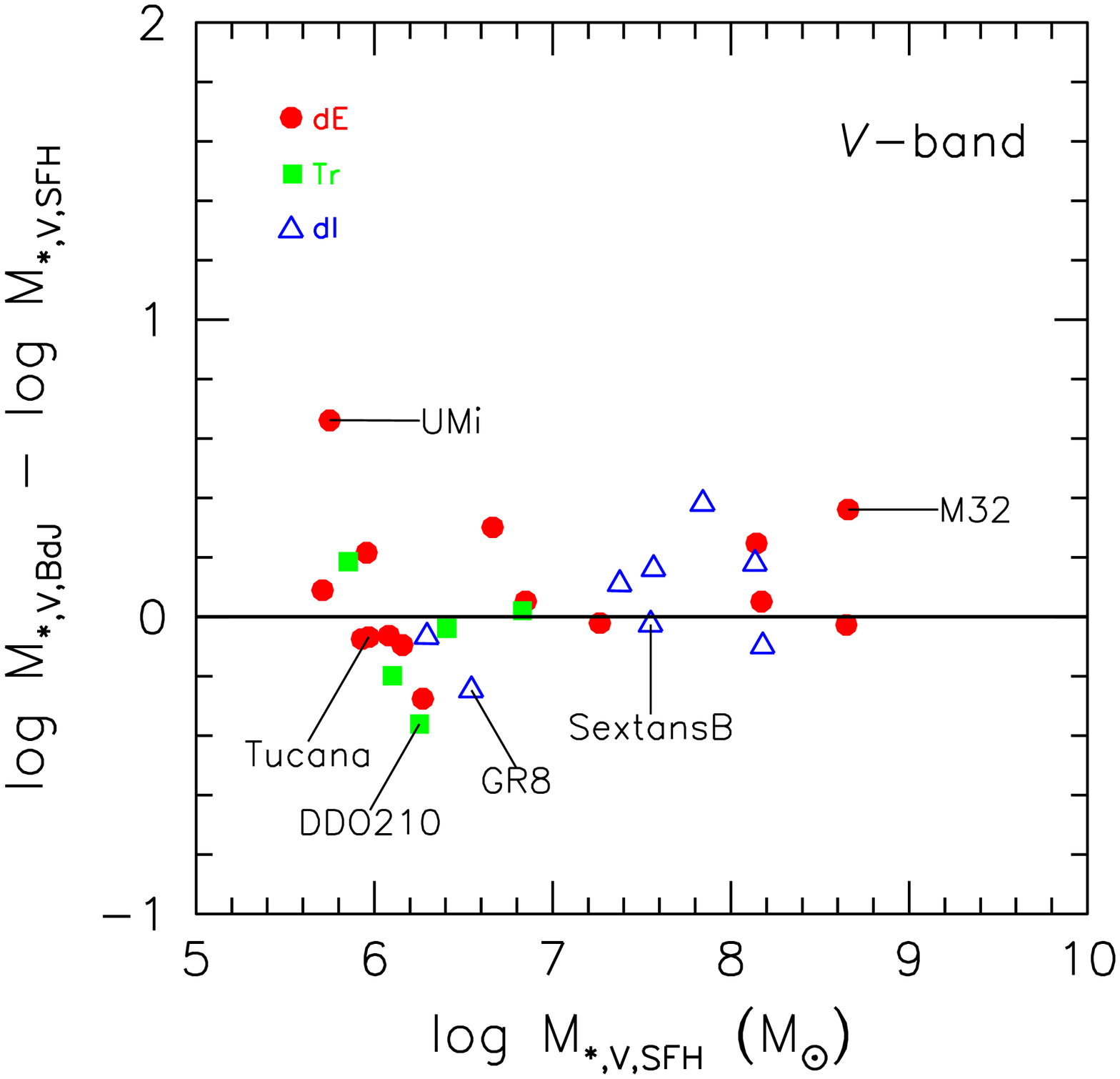}
\caption{Difference in stellar masses computed via BdJ and SFH.}
\label{mbcvsbj}
\end{figure}

The abnormally high $\MSL$ ratio for Ursa Minor dSph (UMi) seen in the 
plot for BdJ values reflects its unusually red colour (see \tab{table}).  
UMi's redness makes it an outlier also in \fig{mbcvsbj}.
However, the $\MSL$ estimated from the SFH approach, which does not
require colour information, yields a reasonable value
in agreement with the observed trend for other Local Group galaxies.

We compare in \Fig{mbcvsbj} the stellar masses computed from the
$\MSL$ ratios of the two methods.  There is good agreement, with 
the exception of UMi.

The stellar masses and stellar mass ratios computed through BdJ 
are generally slightly higher than those computed via SFH.  
This may be explained by the fact that 
BdJ calibrated their colour-$\MSL$ correlation under the assumption of
maximal discs, making the BdJ $\MSL$ ratios upper limits.

\begin{figure*}
\epsscale{2.2}
\plottwo{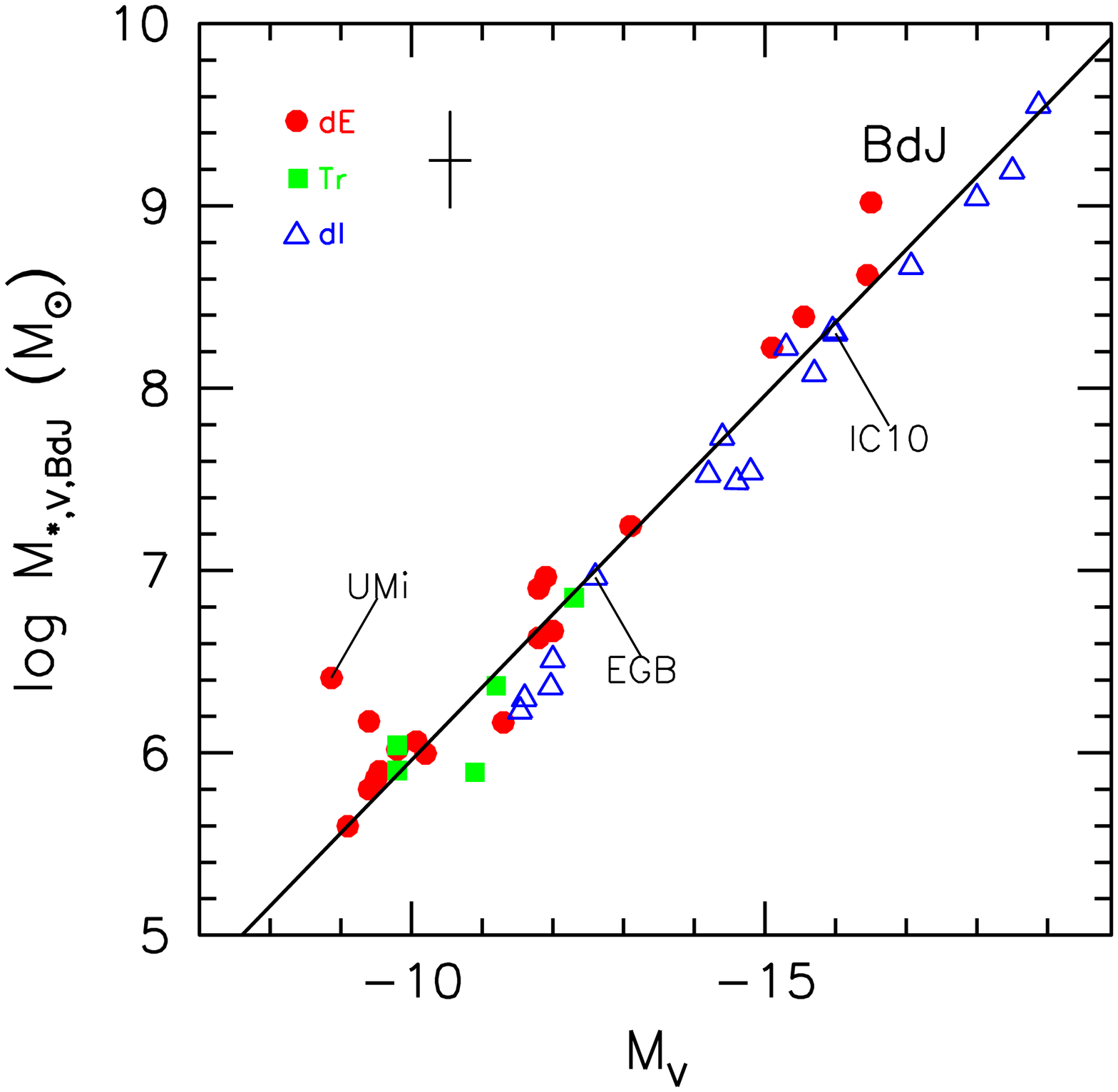}{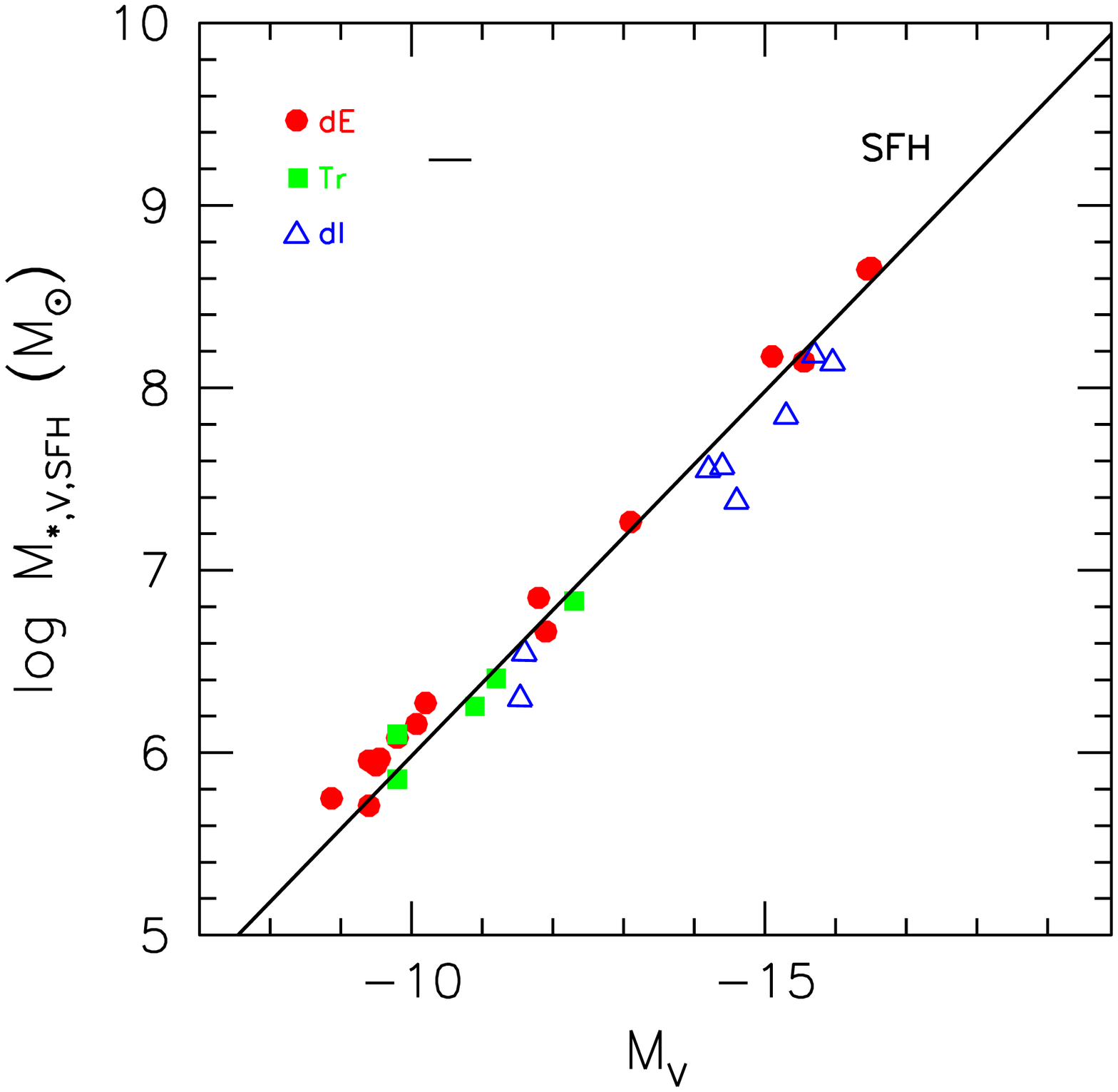}
\caption{
Comparison of stellar masses with the absolute magnitude $M_V$.
The BdJ and SFH computed values are on the left and right respectively.
The solid line has a forced slope of -0.4 and a free vertical offset.
}
\label{mstarm}
\end{figure*}

The distribution of stellar masses, $\Ms$, against absolute $V$ magnitude,
$M_V$, is shown in \fig{mstarm}.  
The solid line has a forced slope of -0.4, 
while the zero-point is adjusted to fit the data.  
Once again, the $\Ms$ values for the dI galaxies 
tend to fall below the dE's as a result of 
the younger populations in dI's.

Both methods of calculating $\MSL$ yield comparable 
values of log $\Ms$ within 3\% (see \tab{avgML}).  Thus, the observed
luminosity scaling  
relations (\eg \ $\Sigma-L$, [Fe/H]-$L$, $R_e$-$L$, [O/H]-$L$, $V$-$L$,
$\sigma-L$) will remain intact when translating them to physical
scaling relations.

\begin{table}
\centering
\begin{minipage}{50mm}
\caption{\small Median $\MSL$ for different galaxy types\label{avgML}}
\begin{scriptsize}
\begin{tabular}{l c c c c c c}
\hline\hline
 & & \mcol{2}{c}{$B$-Band} & & \mcol{2}{c}{$V$-Band} \\
\cline{3-4} \cline{6-7} \\
\colh{}&\colh{}& \colh{SFH} & \colh{BdJ}&\colh{}& \colh{SFH} & \colh{BdJ} \\ 
\hline
dI       &   & \MLBCbi  & \MLBJbi &   & \MLBCvi  & \MLBJvi  \\
Tr       &   & \MLBCbt  & \MLBJbt &   & \MLBCvt  & \MLBJvt  \\
dE       &   & \MLBCbs  & \MLBJbs &   & \MLBCvs  & \MLBJvs  \\
\hline
\end{tabular}
\end{scriptsize}
\end{minipage}
\end{table}

\subsubsection{Choice of Methods}
\label{2data}

Given the incompleteness of our original data bases, we establish various
guidelines in \tab{bdata} to assign final $\MSL$ values per galaxy.

We can use a median $\MSL$ value for all the 
galaxies according to Hubble type, or assign the
$\MSL$ for each galaxy based on the BdJ
or SFH methods.  The
first option recognises the uncertainties of both methods of 
calculating $\MSL$ (namely, the uncertainties in the SFH, 
and the uncertainties in extrapolating BdJ's colour-$\MSL$
correlation down to dwarf galaxies), 
while the second method uses the full available
data to make educated guesses for $\MSL$.  We have thus created 
two data sets of stellar data according to these methods, 
calling the first ``a,'' and the second ``b.''  

For the ``a'' set, we assign the average of the SFR and BdJ median 
values of the $\MSL$ from \tab{avgML}.  
For the ``b'' set, we first shift the BdJ $\MSL$ ratios down so that 
their median values per Hubble type match those from SFH.  We 
then assign their $\MSL$ ratios according to our prescription in \tab{bdata}.

The ``a'' and ``b'' sets are consistent with each other within their
errors, and our choice of ``a'' or
``b'' makes no difference to our conclusions.  
In the subsequent analysis, we employ the ``b'' set since it 
utilises all the available data to calculate $\MSL$, and preserves any
intrinsic scatter.  

\subsection{Surface Density}

The luminosity profiles of the dI and dE galaxies have been
fitted (M98 and references therein)
with an exponential function whose inner extrapolation
provides the central surface brightness $\Sigma_{0}$, which we
translate to a stellar surface density, $\mus$ ($\Msun \kpc^{-2}$)
using the $\MSL$ derived above.

M 32 has a very high central surface brightness of 11.5 mag arcsec$^{-2}$ 
(GGH03),
yielding $\log \mus = 12.1$ ($\Msun ~\kpc^{-2}$).  This is four orders
of magnitude higher than other galaxies of similar luminosity.  
M 32 indeed belongs to the rare class of compact dwarf galaxies
\citep[\eg][and references therein]{zie98,mie05}.
We thus exclude M 32 from analyses involving $\mus$.
However, we retain it for other tests.

\subsection{Velocity}
\label{velsig}

The velocity data for the dwarf galaxies consist of a 
rotational velocity (corrected for inclination),
$\vrot$, and/or a projected central velocity dispersion, $\sigma$.
The dI's typically have a rotation velocity that is larger than
the velocity dispersion (exceptions are Leo A, GR 8 and Sag DIG - see
\citealp{cou07b}),  
whereas dE's only have a measurable velocity dispersion.

We wish to define as a global parameter a characteristic velocity $V$ 
that will represent in both cases the depth of the 
potential well, which for dwarf galaxies is believed to be
dominated by the dark halo. 
The measured rotation velocity in the disky dI's is taken as is,
assuming that it approximates the virial halo velocity
and is a near lower bound to the maximum rotation velocity.
We then ask, what does a dE's velocity dispersion tell us 
about its potential well?  What would have been its rotation velocity
if it had been disky in the same halo with the same stellar mass?  To 
answer these questions, given the tightness of the Tully-Fisher (TF) relation, 
we assume that the halo mass scales with the 
stellar mass in the same way for both dE's and dI's (refer to \sec{stf}) 
and adopt the following 
algorithm:  take $V = {\rm max}\{\vrot,X\sigma\}$ (or the appropriate 
value when only one of $\vrot$ and $\sigma$ is available)
where $X$ is chosen
to minimise the scatter in the TF relation (log $V$ vs. log $\Ms$).
We find that $X=1.92$ which is consistent with the $V$-$\sigma$ 
relation for large disc galaxies \citep{cou07b}, but is larger 
than the $X=\sqrt{2}$ solution for an isothermal sphere.
Had we chosen to minimise the baryonic Tully-Fisher (BTF)
relation instead, we would have found a lower value of $X$ but 
slightly larger scatter than for the TF relation for these dwarf 
galaxies (see also \sec{stf} for further motivation to minimize 
the TF instead of the BTF relation).
The choice of the isothermal solution $X=\sqrt{2}$ would yield 
a TF relation that is slightly steeper (in log $V$ vs. log $\Ms$), 
but the galaxy types still form a single relation.  Furthermore,
the results of our PCA analyses (described in \sec{tools}) are 
statistically identical for both $X=\sqrt{2}$ and $X=1.92$. 

In summary, our algorithm sets
$V=X\sigma$ for the dE galaxies, and $V=\vrot$ for the dI galaxies (except 
for GR 8, Leo A and Sag DIG where $\sigma$ is greater than the
maximum $\vrot$ of the ISM).

\subsection{Radius, Baryon Mass, Metallicity and SFR}

We use the exponential scale length scaled with the distance
estimates provided in M98 as the characteristic stellar radius 
of galaxies.  We also use the \HI mass provided by GGH03 and
our derived stellar mass to estimate a baryonic mass.  For most 
of the dE's the quoted \HI mass is an upper limit.

Iron abundances [Fe/H] are taken from GGH03, which
are based on spectroscopic and photometric 
measurements of red giant branch stars.  GGH03 use the mean 
abundances of the old stellar populations which can be consistently 
measured across the galaxy types.  Oxygen abundances are
available only for galaxies with current star formation, so besides 
the dI's, 
only four dE's and one Tr galaxy have available oxygen data. 
Thus, we use the iron abundances to trace the total metallicty $Z$
and assume a linear relation between them.  Recalling that [Fe/H] is 
logarithmic, we then define $Z$ as:
\be
\log (Z/Z_\odot) = {\rm [Fe/H]}
\label{metal}
\ee
where we adopt $Z_\odot=0.019$ \citep{and89,gir00,car01}.

The current star formation rate (SFR or $\sfr$) data for the dI galaxies, 
estimated 
from observed H$\alpha$ flux, are taken directly from 
M98 or vdB00.  dE galaxies
have very little or no current star formation.

\begin{table*}
\begin{minipage}{116mm}
\caption{\small $\MSL$ assignments for data set ``b''\label{bdata}}
\begin{scriptsize}
\begin{tabular}{@{}llll@{}}
\hline\hline
\colh{} & \colh{Case 1} & \colh{Case 2} & \colh{Case 3}     \\
\hline 
dI & average $\MSL$ ratios from BdJ and SFH & whichever datum exists & $\MSL$ from data set ``a'' \\
dE & $\MSL$ from SFH                        & whichever datum exists & $\MSL$ from data set ``a'' \\
Tr & $\MSL$ from SFH & &    \\
\hline
\end{tabular}
\end{scriptsize}
{\scriptsize Note. ---
The cases are: 
1. enough data exist to calculate $\MSL$ from both SFH and BdJ
methods;  
2. data exist for only one method of computing $\MSL$; 
3. not enough data exist for either SFH or BdJ methods. (For Tr
galaxies, only Case 1 applies.)
}
\end{minipage}
\end{table*}

\subsection{Error bars}

Due to the uncertainties in the determination of $\MSL$ and
the heterogeneous data base, error bars are not taken into 
account in the subsequent analysis.  However the typical
error bar size relative to the scatter in distribution of 
the galaxies is shown in the upper left corner of all figures.  
The estimated error bars for $\log \Ms$ and $\log \mus$ reflect 
the difference between the SFH and BdJ calculations $\MSL$, 
and are on average \erangeMs\% and \erangemus\% 
of their respective ranges.  The 
error bars are typically \erangeV\% for $\log V$ and \erangeZ\%
for $\log Z$ (Eva Grebel, private communication).  
No estimates are provided for the uncertainty in SFR.

\section{Methods of Analysis}
\label{tools}

Throughout this analysis, we study the relations between the dwarf
galaxy structural and SF parameters using linear regression and
principal component analysis (PCA).

For the two-dimensional analyses, we investigate the correlation in 
log space between pairs of the structural and SF quantities.  
We plot the distribution of each pair, obtain the best-fitting line 
by the method of bisector least-squares,
and calculate the standard Pearson linear correlation 
coefficient $r$ according to \cite{pre92}.

We use PCA as a tool to quantify the shape of the distribution of the
galaxies in parameter space.  
Given a data set with $m$ parameters (such as stellar mass, metallicity, etc), 
the PCA quantifies the distribution of the data 
along orthogonal basis vectors of the 
parameter space.  PCA outputs the vectors {\bf V}$_k, k=1,m$ and their
eigenvalues $D_k$.  
Appendix B describes the derivation of {\bf V}$_k$
and $D_k$ (or $D_{k,k}$ as in the Appendix).

The following is an interpretation of the eigenvectors and eigenvalues:

The eigenvectors, which are orthonormal, lie in the directions of greatest
variance in the data.  The ratios of the eigenvalues are a measure of the 
relative strength of the variance along each corresponding eigenvector.  
For example, if the distribution of the data were a line in a parameter 
space of 3 parameters, the vector {\bf V}$_1$ with the largest eigenvalue $D_1$
will lie along that line, while the other two vectors will be
orthogonal to it.   
Furthermore, the eigenvalue $D_1$ will be much larger than the 
other two eigenvalues, which will be similar in size, 
reflecting the extended 
distribution of the data along {\bf V}$_1$, and relatively little 
extension in the other two directions.  Such a distribution is characterised by one 
primary parameter.  If the distribution of the data
is a plane in 3-D space, two of the eigenvectors will lie in the 
plane and will have comparably large eigenvalues, while the third vector, 
lying orthogonal to the planar distribution will have a relatively small 
eigenvalue.  Such a distribution is characterised by two primary parameters.

In fact, for 3-D space, we can more generally associate
a ``linear'' distribution with ``prolate'' ($D_1$:$D_2$ $\gg$ $D_2$:$D_3$), 
and a ``planar'' distribution with ``oblate'' ($D_1$:$D_2$ $\ll$ $D_2$:$D_3$).

In order to meaningfully quantify the shape of the data distribution,
we standardise the data before performing the PCA.  Standardisation
divides the data set by the standard deviation of the parameters, so
that the eigenvalues are {\it independent} of parameter range.
In other words, standardisation gives us a shape for the
distribution with the ranges of all the axes on equal footing.
Without standardisation the shape of the distribution will change 
from being linear to planar, depending on their relative scalings.  
In fact, the scaling relations themselves can be derived directly from the
elements of the first eigenvector of the unstandardised analysis.
On the other hand, standardisation 
will output eigenvalues that are independent of the scalings, and therefore 
insensitive to whether we choose for instance to use $V^2$ instead of $V$.
Therefore, throughout this study, we present the results
of standardised PCA along with the standard deviations of each
parameter.  (The scaling relations may be recovered by dividing the
components of the first principal eigenvector {\bf V}$_1$ by the
corresponding standard deviations.)

 We caution the reader that standardisation assumes a complete, 
 unbiased representation of all galaxy types for all considered 
 parameters.  Given the depth of current imaging surveys, we suspect 
 that only much fainter LG dwarfs are truly missing, and our assumption 
 of standardisation is reasonable.  Performing our analysis using  
 strictly unstandardised PCA would yield tighter fundamental lines   
 in every case due to the large range of $\Ms$ relative to the  
 other parameters.  Standardisation is nonetheless preferred for  
 the reasons stated above. 

In the following sections, we present our 2D and PCA analyses first
for the structural parameters, and then for the SF
parameters.  

\section{Results: The Structural Parameters}
\label{stranalysis}

\begin{figure*}
\epsscale{2.2}
\plottwo{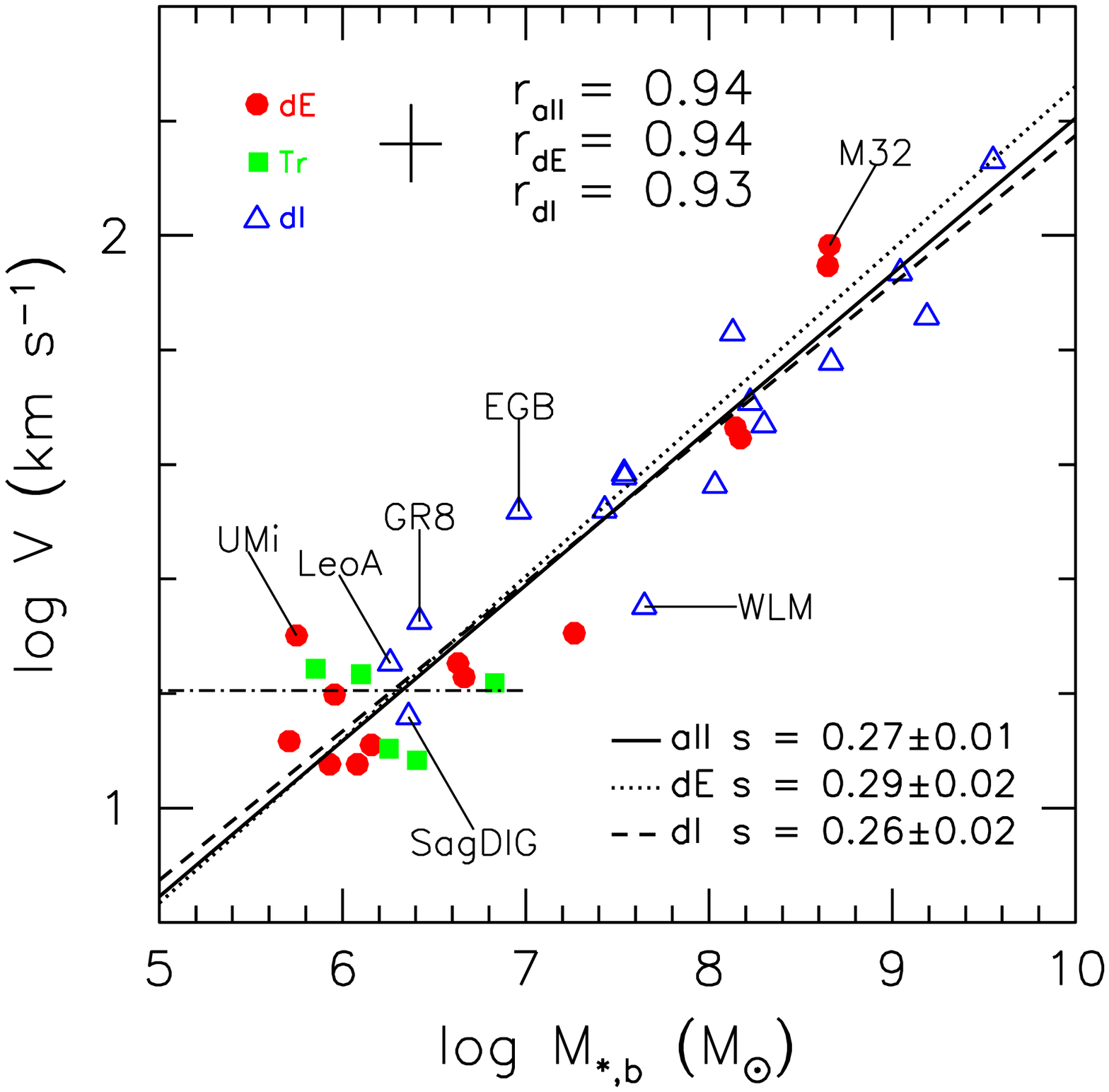}{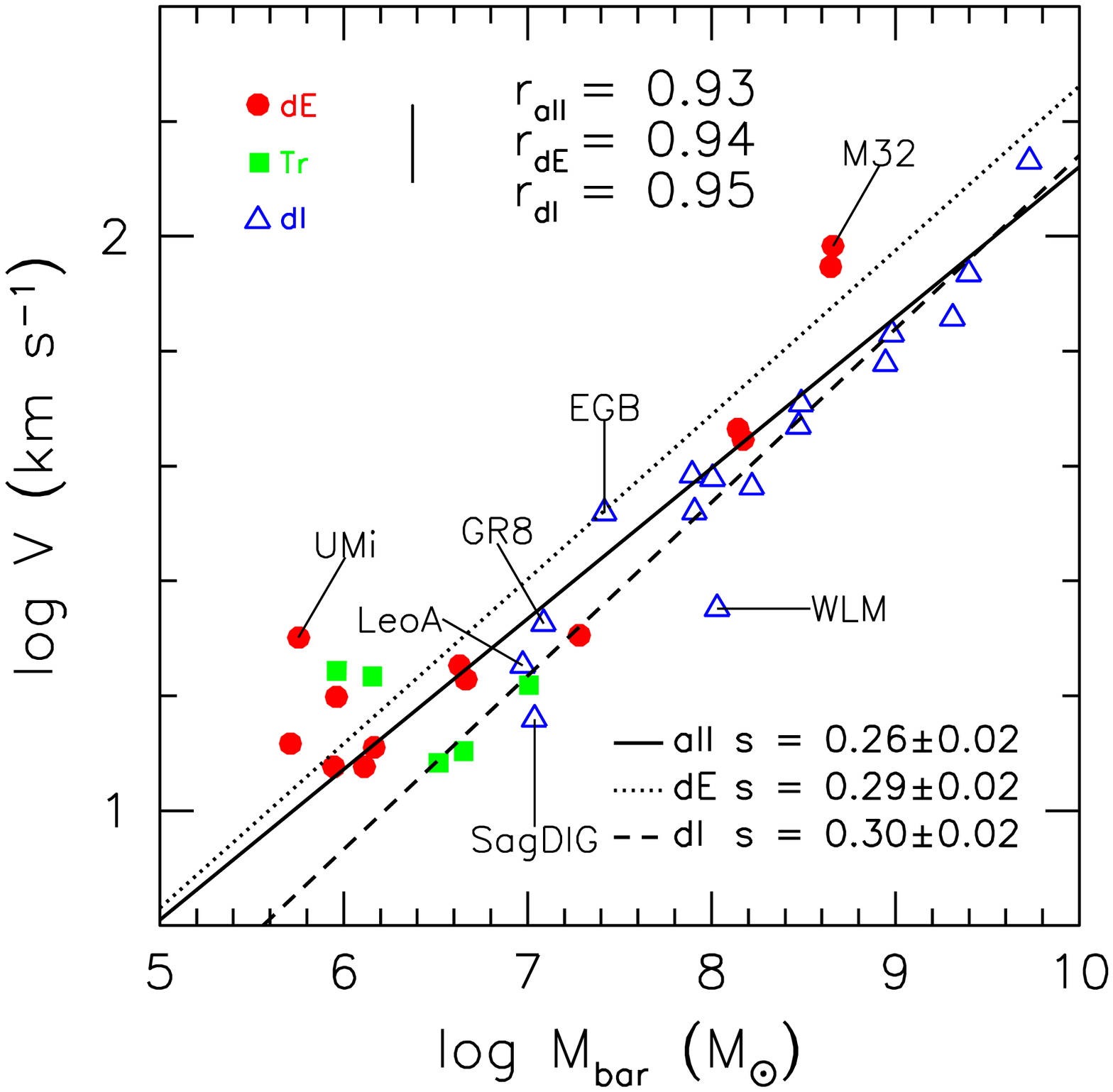}
\caption{Left: Circular velocity vs. stellar mass for LG dwarf
  galaxies.  Right: Circular velocity vs. ``baryonic mass'' ($\Mbar = \Ms +
  \MHI$).
We show bisector fits for all the galaxies
(solid), and separately for the dE (dotted) and dI (dashed) types.  
The dash-dotted line delineates galaxies below $\Ms=10^{6.9}\Msun$. 
The Pearson correlation coefficient $r$ is shown at the top and the 
bisector fit slope $s$ at the bottom of each figure.
}
\label{tf}
\end{figure*}

\subsection{2D: Velocity and Mass}
\label{stf}

\Fig{tf} shows the correlation of characteristic velocity versus $\Ms$ for
LG dwarf galaxies with $V \prop \Ms^{\scV}$ and $r=\rV$.  
Recall that $V$ is constructed from $\vrot$ and $\sigma$ in such a way as
to minimise the scatter in this scaling relation.  So naturally, the dI and
dE galaxy types seem to lie on the same relation with uniform spread.
However, had we used $V=(\sqrt{2}$ or $\sqrt{3})\sigma$ instead of $\sim 2\sigma$
the dI's and dE's would still appear to lie on the same relation.
Note also that the smallest dwarfs cluster just above $V \sim 10$km s$^{-1}$
and could even be described by a constant $V$ (dash-dotted line) which 
\citet{dek03} explain might be
due to radiative feedback.

The LG dwarf galaxies fall below $V_c \lsim 90 \kms$.  
The $V-\Ms$ relation of LG dI's is consistent with that 
of \citet{mcg05}, with a slope of $\scVdi$ or 
$\Ms \prop V^{\scVdirev}$, which is not significantly different 
from $\Ms \prop V^{4}$.

Adding the gaseous mass from GGH03 ($M_{bar} = \Ms + \MHI$) 
yields a baryonic Tully-Fisher (BTF) slope for the LG dI's that is 
significantly shallower ($\Ms \prop V^{\BTFdIrev}$) 
than that reported by \citep{mcg05}, but consistent with the BTF
relations of both BdJ and \cite{geh06} (\fig{tf} left).  We also find that
for dI's with log $V$ $>$ 1.5, we find that the scatter in the LG BTF 
relation is reduced significantly compared to \fig{tf} (right).  
These dI's lie almost precisely along the BTF relation found by 
\citep{mcg05}, with comparable scatter.  \cite{mcg05}
describes a break in the distribution of circular velocity
versus stellar mass (or total light) for galaxies with $V_c \lsim 90
\kms$.  This break however disappears if $V_c$ is plotted against
total (luminous + gaseous) mass.  

The shallower overall slope of the dI BTF relation is likely due to
the different prescriptions for calculating $\MSL$.  \cite{mcg05} used
an optimal empirical relation (see their section 3.3), while
we used stellar population models, as did BdJ (see \sec{stellarmass}).
\cite{mcg05} showed that this difference in methods is enough to account 
for the shallower BTF slope found by BdJ (which is consistent with ours).

The contribution of the \HI mass to dE galaxies is negligible 
and we find two distinctly offset BTF relations for dI and dE
galaxies.  (This is true even had we used $V=\sqrt{2}\sigma$ for the
dE's.)  This offset would have
been larger had we used the suggestion of \citet{pfe05} that 
$M_{bar} = \Ms + 3 \MHI$.  This suggests that although the BTF
may more fundamentally describe the kinematics of dI's, the stellar 
relation (\ie the TF) 
may be a more fundamental description of the entire
dwarf population (since the same TF describes all galaxy types).
(Recall also that had we minimised the scatter of the BTF in determining 
the $X$ of \sec{velsig}, the scatter of the BTF would have been slightly 
larger than the minimised scatter of the TF.)
Using baryon mass instead of stellar mass
also increases the offset 
between the types in the $\Rs-\Mbar$, $\mus-\Mbar$, and $Z-\Mbar$
relations.

\begin{figure}
\epsscale{1}
\plotone{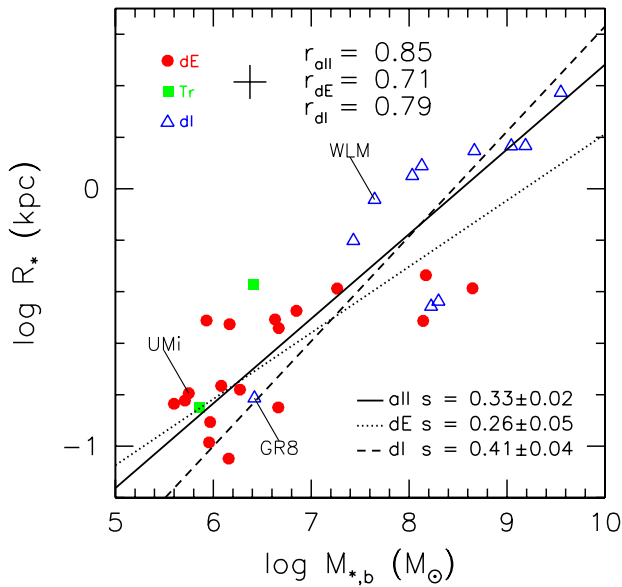}
\caption{Exponential radius vs. stellar mass.
We plot the two-dimensional linear regression on all the galaxies
(solid), and separate fits to the dE (dotted) and dI (dashed) types.  
}
\label{Mrp}
\end{figure}

\subsection{2D: Radius and Mass}
\label{sMrp}

 \Fig{Mrp} shows the correlation between exponential radius 
 and stellar mass fitted as $\Rs \prop \Ms^{\scrp}$, with $r = \rMsrp$. 
 The galaxy types populate different regimes of the relation, with  
 dI galaxies being bigger and more massive than dE's, yet they are  
 both described by the same $\Rs-\Ms$ relation.  This is interesting  
 in itself as it may indicate a similar physical origin to the relation, 
 (such as halo mass), which may simultaneously be responsible for 
 their differences. 
 
\cite{she03} have plotted the median of the distribution of the sizes 
of the SDSS galaxies as a function of stellar mass (see their Fig. 11).  
The sizes are measured for the $z$-band and are given as Sersic
half-light radii.  If we neglect bandpass differences, we can make 
a rough comparison of the LG data with SDSS data.  Considering that 
the exponential half-light radius is $R_{1/2}=1.68\Rs$, we find that
the LG dE's appear to be the low-mass extensions of the early-type 
distributions of sizes for the SDSS galaxies.  The SDSS early-type 
galaxies follow the relation $\Rs \prop \Ms^{0.55}$ while LG dE's 
have a shallower $\Rs \prop \Ms^{\scrpde}$.  This shallower slope is
consistent with the lower end of the size-luminosity relation of
\cite{gra08} while we extend their relation
to lower masses.

However, while the LG dI's overlap the $\Rs-\Ms$ relation of the 
SDSS late-type galaxies, they extend the relation which a  
steeper slope of $\scrpdi$ compared to the slope of $0.15$ at 
the low-mass end of the SDSS sample. 

We can derive a stellar size-mass, $\Rs-\Ms$, relation for large 
spiral galaxies from the compilation of scaling parameters for 1300 
local disc galaxies by \cite{cou07}.  We use the 
$I$-band exponential scale radii and luminosities, $V-I$ colours, 
and colour-$\MSL$ relations of BdJ \citep[see ][]{dut07} to
derive a $\Rs-\Ms$ relation for large galaxies.  We find 
that the dI's overlap with the $\Rs-\Ms$ relation of bigger 
disk galaxies and with comparable slope.

\subsection{2D: Surface Mass Density and Mass}
\label{sMmu}

\begin{figure}
\epsscale{1.0}
\plotone{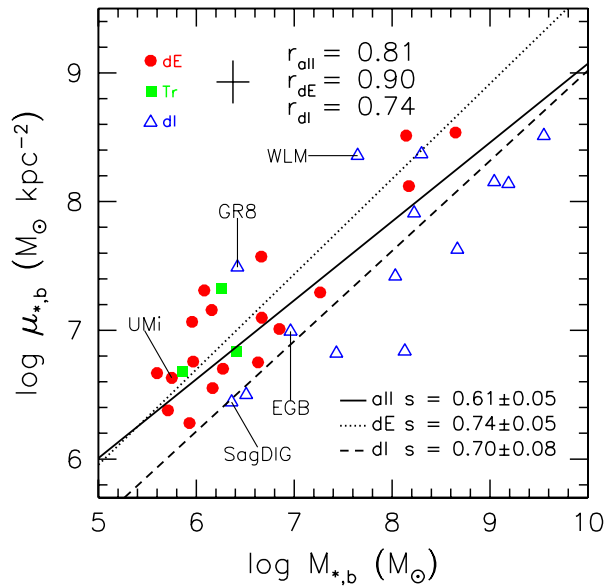}
\caption{Central stellar surface density vs. stellar mass.  
We plot the two-dimensional linear regression line 
of all the dwarf galaxies (solid), and separate regressions for 
the dE (dotted) and dI (dashed) types.
}
\label{Mmu}
\end{figure}

The dependence of radius and stellar mass is also expressed via the
relation for stellar surface density ($\mus \prop \Ms/\Rs^2$) and
stellar mass.  \Fig{Mmu} shows a correlation between 
$\mus$ and $\Ms$ that spans
\decMs decades in $\Ms$ and \decmus decades in $\mus$. 
The best-fitting relation is 
$\mus \prop \Ms^{\scmus}$ with $r=\rmus$.

With the exception of WLM and GR 8, dI galaxies
have a lower mean surface mass density than dE's.  Since
both $\Ms$ and $\mus$ depend on $\MSL$ in the same way, 
plotting $\mus$ against $\Ms$ or $L$, as previously done by 
GGH03, yields identical offsets between the dE and dI types. 
These best fit lines have the same slope of $\scmusde$, but are offset 
by about \musoff dex from the centre of the distribution.

\citet[see their Fig. 7]{kau03b} showed an analogous distribution in the 
plane of surface density (within the half-light radius) 
versus stellar mass for 123,000 SDSS galaxies
with $\Ms > 10^8\msun$.  They find a weak systematic dependence of 
about $\mus \prop \Ms^{0.2}$ (our eye-ball estimate) for the HSBs and 
$\mus \prop \Ms^{0.63}$ (their quote) in the LSB regime 
with $10^8 < \Ms < 3 \times 10^{10} \msun$.  After considering the $\mus-\mu_{e}$
conversion, the LG dwarf galaxies
extend this relation down to $8 \times 10^5 \Msun$ and together, their
$\mus-\Ms$ relation is consistent with 
$\mus \prop \Ms^{0.6}$.  

\cite{she03} showed the median of the same distribution, but separated
into late and early-type distributions.  They show that the
early-types tend to have higher surface density, 
consistent with our finding for LG galaxies.  In fact, the LG
dE's and dI's again seem to extend early and
late-type surface density relations to lower masses.  (Our eye-ball
estimate of their slopes: $\sim0.65$ for both late- and early-types, 
compared to $s=0.61 \pm 0.05$ for LG dwarfs).

\begin{figure}
\epsscale{1.0}
\plotone{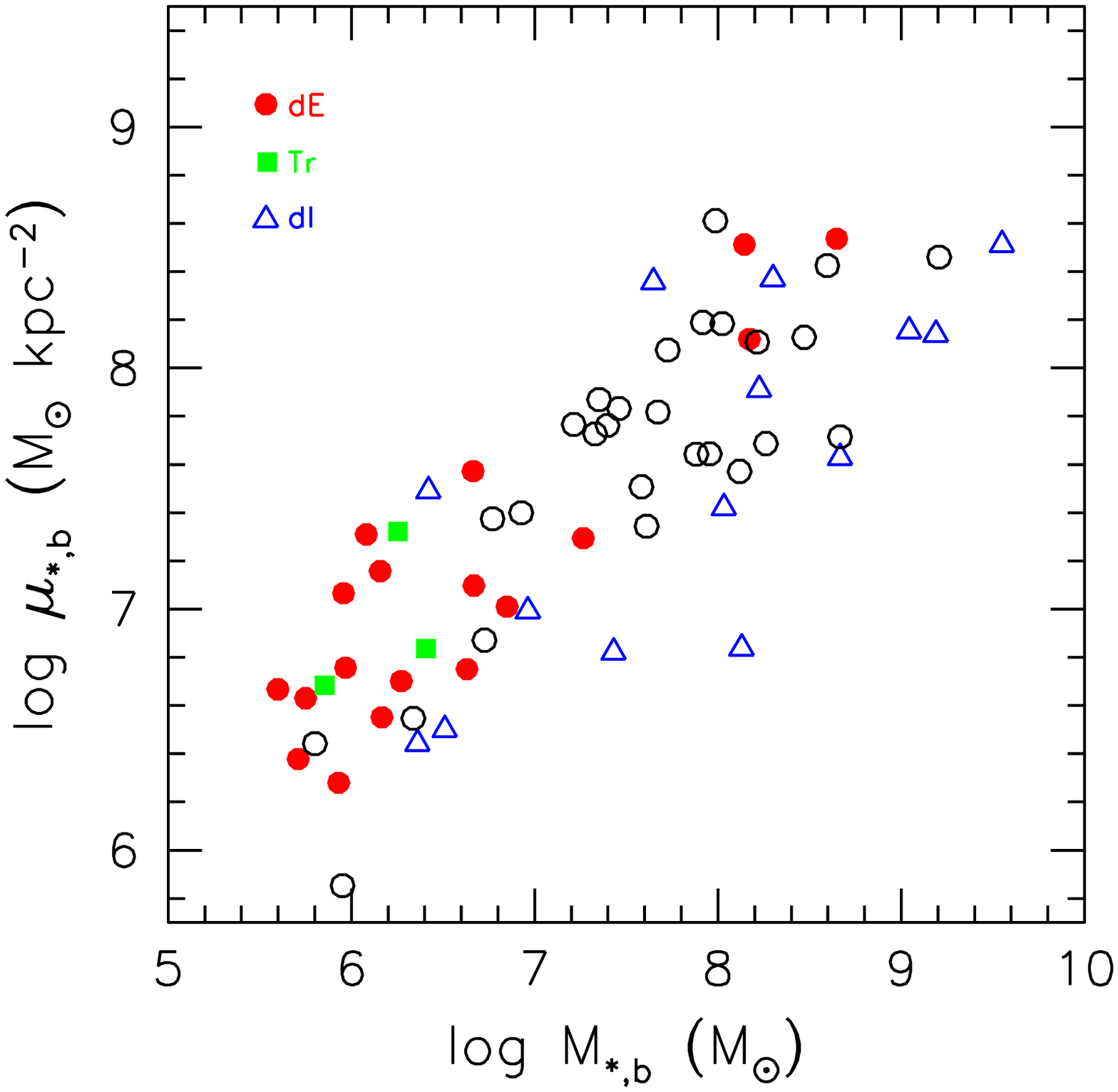}
\caption{The $\Ms$-$\mus$ relation of the LG dwarf galaxies plotted
  together with the dI's of \citet{vad05b} (open circles).
}
\label{compareMmu}
\end{figure}

\cite{vad05b} studied a sample of 34 dI galaxies and provided central surface
brightnesses and $K$ and $J$-band magnitudes for these galaxies using
a hyperbolic secant (sech) function to fit their surface brightness profiles.
Using the colour-$\MSL$ relations of \cite{bel01}, we converted their
sech magnitudes and central surface brightnesses to stellar
quantities, and compared their $\Ms$-$\mus$ relation to that of the LG
dwarf galaxies.  We find that the $\Ms$-$\mus$ relation  
is significantly tighter ($r = 0.86$) than 
their magnitude relation ($M_{K,{\rm sech}}$-$\mu_{0,{\rm sech}}$, $r = 0.71$).  
Secondly, we find that their dI's lie neatly along the same
$\Ms$-$\mus$ relation as the LG dwarf galaxies, even though the
surface brightness profiles of the LG dwarfs were fit to exponential
functions.  We plot the data from \cite{vad05b}, which we converted to
stellar quantities, together with the LG data in \fig{compareMmu}.

\subsection{PCA: The Fundamental Line in Structural Space}

We use PCA as outlined in \sec{tools} to quantify the shape of the 
distribution of the LG dwarf galaxies in the (log) space of the 
structural parameters ($\Ms$, $\Rs$, $V$).  We present the results 
graphically in \fig{strucfig} and quantitatively in \tab{strucpca}.

\Fig{strucfig} shows the axes {\bf Y}$_j$ (defined in Appendix B)
as the coordinates of the data vectors projected onto 
{\bf V}$_j$, for $j=1...3$.  Each panel represents different views
along the three dimensions of the distribution.
On visual inspection, the shape of the distribution is linear 
in \fig{strucfig}.  

We list the components of the three {\bf V}$_j$ vectors in \tab{strucpca}.
The first vector points in the direction
of the fundamental line in this scaled basis.  The other two vectors point 
in the directions orthogonal to the line.  
By construction, the components of the first principal vector {\bf V}$_1$ 
are $\sim 1/\sqrt{n}$ where $n$ is the number of dimensions (in this case 3) 
since standardisation places all the data along a diagonal.
The vector components are in the basis of 
the original four parameters scaled by their standard deviations, from 
which the scaling relations are derived.
(In the unstandardised analysis, the scaling relations are recovered 
directly from the components of the first eigenvector.)
For example, if the first eigenvector is {\bf V}$_1$, and its $j$'th component 
is {\bf V}$_1$($j$), then the scaling relation between the parameters 
log $\Ms$ and log $V$ is 
\be
\frac{ {\bf V}_1(\Ms) / \sigma_{\Ms} }{ {\bf V}_1(V)  / \sigma_{V} } = 
\frac{0.60/1.22}{0.58/0.32}
\ee 
yielding $\log V \prop (0.27 \pm 0.01)\log \Ms$
which is consistent with the 2-D fit in \sec{stf}.  The scaling
relation between log $\Ms$ and log $\Rs$ is listed in the note 
under \tab{strucpca},
and is also consistent with the 2-D fit in \sec{sMrp}.  

The eigenvalues associated with the principal vectors quantify the
shape of the  
distribution of the data.  The eigenvalue associated with the 
first principal vector is $90.1 \%$ of the sum of all the eigenvalues.
The strength of the second eigenvalue is only $8.4 \%$.  The
variance along the first principal vector is more than $10$ times 
greater than the variance along the second principal vector.  
The ratio of the second-to-third eigenvalues is only $5.6$.
In other words, the extensions of the data along the second and 
third principal vectors are more similar to each other
than they are to the extension of the data along the first principal vector.
This describes a linear or prolate distribution, according to the
criteria in \sec{tools}, and is characterised by one primary parameter.

The standard method of displaying fundamental
lines or planes (edge-on) from PCA is to plot one parameter against a linear
combination of the other parameters.  So the LG data would be
plotted as $\logMstarv_b$ versus
\be
(2.92 \pm 0.31)\logv4+(0.72 \pm 0.32)\logrp+(3.35 \pm 0.23)
\ee
from the third principal vector.  
This is equivalent to displaying only the upper-left panel in
\fig{strucfig}, but rotated so that the log $\Ms$ lies along the
$y$-axis, and scaled so that the data lie on the $y=x$ line. 
Such a presentation of
the data, with its confusing mix of parameters on its $x$-axis, 
is not spatially intuitive as it fails to display their
distribution from other angles.  We feel that our method of displaying 
the output of PCA is more useful since 
the data are displayed along all the principal axes of their distribution,
giving the viewer a more intuitive feel for its shape.

\begin{figure}
\epsscale{1}
\plotone{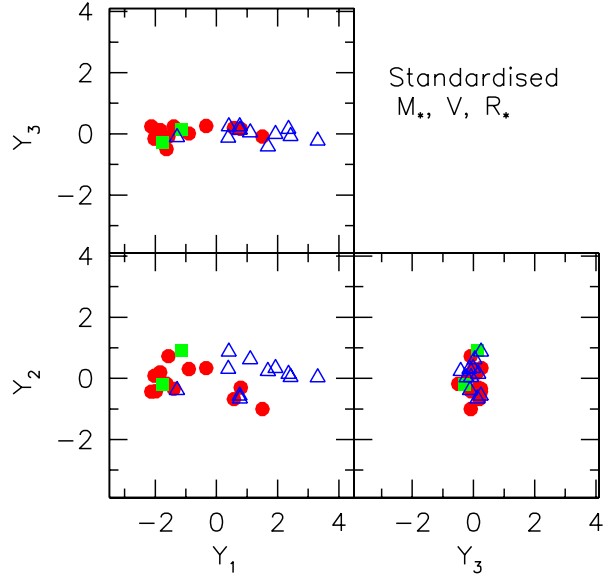}
\caption{The fundamental line in structural space.  
The axes lie in the directions of the {\bf V}$_j$ eigenvectors defined 
in the text and the data are projected onto them.  The components of the 
eigenvectors are given in \Tab{strucpca}.
}
\label{strucfig}
\end{figure}

\begin{table}
\centering
\begin{minipage}{84mm}
\setlength{\tabcolsep}{1.0mm}
\caption{\small PCA results in structural space\label{strucpca}}
\begin{scriptsize}
\begin{tabular}{ r r r r r l}
\hline\hline
\colh{$X$}     & \colh{$\logMstarv_b$} & \colh{$\logv4$} & \colh{$\logrp$} & \colh{Eigen-} & \colh{Ratios}\\
 & & & & \colh{values (\%)}& \\
\hline
{\bf V}$_1$   &  0.60 $\pm$  0.01  &  0.58 $\pm$  0.01  &  0.56 $\pm$  0.01  & 90.1 $\pm$  3.2 & \mrow{2}{*}{\}10.7$\pm$ 3.9 } \\     
{\bf V}$_2$   & -0.23 $\pm$  0.08  & -0.54 $\pm$  0.07  &  0.81 $\pm$  0.03  &  8.4 $\pm$  3.0 & \mrow{2}{*}{\}5.6$\pm$ 2.7 } \\     
{\bf V}$_3$   &  0.77 $\pm$  0.02  & -0.61 $\pm$  0.06  & -0.19 $\pm$  0.08  &  1.5 $\pm$  0.5 &  \\     
$\sigma_X$ &\mcol{1}{c}{ 1.22} & \mcol{1}{c}{ 0.32} & \mcol{1}{c}{ 0.41} &  \\[5pt]
\hline 
\end{tabular} 
\end{scriptsize} 
{\scriptsize Note. ---
The scaling relations from the principal vector: \\
$\logv4$ $\propto$ (0.27 $\pm$ 0.01)$\logMstarv_b$, $\logrp$ $\propto$ (0.36 $\pm$ 0.01)$\logMstarv_b$. \\
 Standard projection: \\
$\logMstarv_b$=(2.92 $\pm$ 0.31)$\logv4$+(0.72 $\pm$ 0.32)$\logrp$+(3.35 $\pm$ 0.23). \\
}
\end{minipage}
\end{table}

\begin{figure}
\epsscale{1}
\plotone{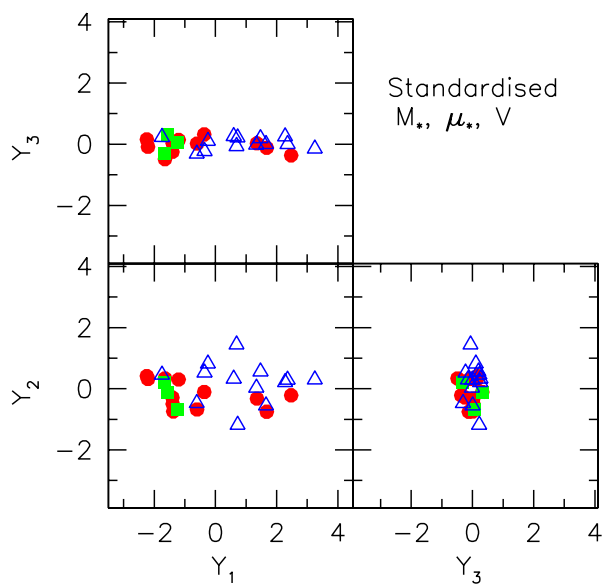}
\caption{Projections of the LG dwarf data in an alternate structural space.  
The components of the PCA eigenvectors are given in \Tab{altstrucpca}.
}
\label{altstruc}
\end{figure}

We also performed PCA on an ``alternate'' structural
parameter space where we have replaced $\Rs$ with $\mus$ in the above
structural space.  We summarise the results
in \fig{altstruc} and \tab{altstrucpca}.  The distribution of the
galaxies is triaxial in this space since the ratios of the
eigenvalues (listed in the table) are comparable.  However, since the first
eigenvalue is at least ($8.0$) times greater than the other
eigenvalues, meaning that the galaxies are at least $8$ times more
extended along one axis than the along the other axes, we say that the galaxies
lie in a linear distribution, characterised by one primary parameter, in this 
alternate structural space.

\begin{table}
\centering
\begin{minipage}{84mm}
\setlength{\tabcolsep}{1.0mm}
\caption{\small PCA results in alternate structural space\label{altstrucpca}}
\begin{scriptsize}
\begin{tabular}{ r r r r r l}
\hline\hline
\colh{$X$}     & \colh{$\logMstarv_b$} & \colh{$\mustarv_b$} & \colh{$\logv4$} & \colh{Eigen-} & \colh{Ratios} \\
 & & & & \colh{values (\%)}& \\
\hline
{\bf V}$_1$   &  0.60 $\pm$  0.01  &  0.54 $\pm$  0.02  &  0.58 $\pm$  0.01  & 87.5 $\pm$  3.8 & \mrow{2}{*}{\}8.0$\pm$ 2.7}\\     
{\bf V}$_2$   &  0.24 $\pm$  0.06  & -0.82 $\pm$  0.02  &  0.51 $\pm$  0.05  & 11.0 $\pm$  3.7 & \mrow{2}{*}{\}7.0$\pm$ 3.2}\\     
{\bf V}$_3$   &  0.76 $\pm$  0.02  & -0.17 $\pm$  0.07  & -0.63 $\pm$  0.04  &  1.6 $\pm$  0.5 & \\     
$\sigma_X$ &\mcol{1}{c}{ 1.19} & \mcol{1}{c}{ 0.71} & \mcol{1}{c}{ 0.32} &  \\[5pt]
\hline 
\end{tabular} 
\end{scriptsize} 
{\scriptsize Note. ---
The scaling relations from the principal vector: \\
$\mustarv_b$ $\propto$ (0.66 $\pm$ 0.03)$\logMstarv_b$, $\logv4$ $\propto$ (0.27 $\pm$ 0.01)$\logMstarv_b$. \\
 Standard projection: \\
$\logMstarv_b$=(0.35 $\pm$ 0.15)$\mustarv_b$+(3.00 $\pm$ 0.24)$\logv4$+(0.28 $\pm$ 0.37). \\
}
\end{minipage}
\end{table}

\section{Results: The Star Formation (SF) Parameters}
\label{speanalysis}

\subsection{2D: Metallicity and Mass}
\label{sMZ}

\begin{figure}
\epsscale{1}
\plotone{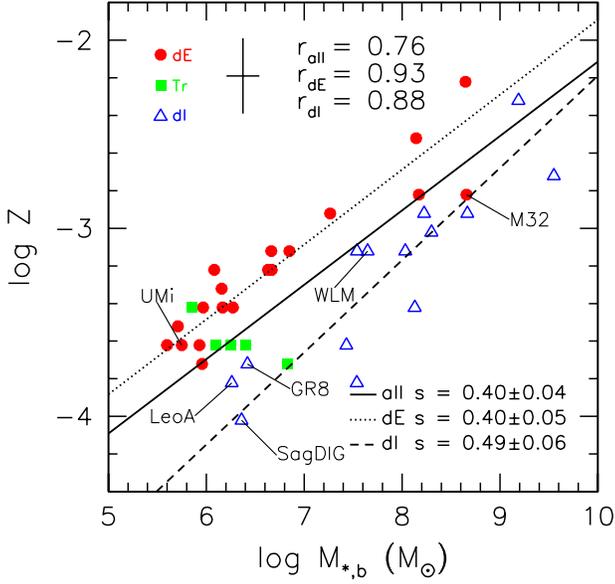}
\caption{Metallicity vs. stellar mass.
We plot the two-dimensional linear regression over
all the dwarf galaxies (solid), and separate fits for dE (dotted), 
and dI (dashed) galaxies.}
\label{MZ}
\end{figure}

\Fig{MZ} shows metallicity versus $\Ms$.  The best-fitting
global scaling relation is $Z \prop \Ms^{\scZ}$ with $r = \rMZ$.
The higher metallicity of dE galaxies relative to the dI galaxies
of the same luminosity has been known for some time (see for example M98), 
and this offset persists even 
for old stellar populations (GGH03).  Not surprisingly, we see the
same phenomenon when using $\Ms$ instead of luminosity.
Their respective correlations are much stronger when fitted separately
(dE's: $\rMZs$; dI's: $\rMZi$).

For SDSS data, \citet{kau03b} reported $Z \prop \Ms^{0.4-0.5}$ 
(depending on the metallicity tracer) for $\Ms < 3 \times 10^{10} \Msun$, 
at the bright end of the dwarf galaxy regime \citep{kau03b}.  This 
is consistent with our global and type-dependent slopes.

\begin{figure}
\epsscale{1.0}
\plotone{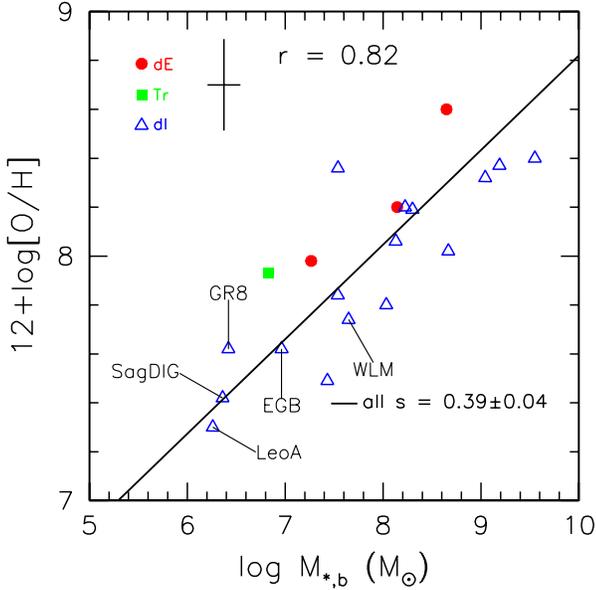}
\caption{Oxygen abundances of the LG dwarf galaxies vs. stellar mass.}
\label{MOx}
\end{figure}

\citet{tre04} showed more specifically that SDSS star-forming galaxies 
follow the relation
12 + log [O/H] $\simeq$ 0.35 log $\Ms$ for $\Ms < 3 \times 10^{10} \Msun$, 
with an overall saturation beyond this transition mass (see their Fig. 6).  
This translates to $Z \prop \Ms^{0.35}$ at the low-mass end 
assuming that the oxygen abundances trace $Z$ linearly.
Using oxygen abundance data from M98, 
we find that the LG galaxies follow 
12 + log [O/H] $\simeq \scOx$ log $\Ms$, consistent with the slope of
low mass SDSS galaxies (\fig{MOx}).  Our slope is also a slightly better match
to the low-mass end of the SDSS galaxies than that of \cite{lee06}.

However we find that the LG dwarf galaxies 
have oxygen abundances that are on average 0.3 dex lower than
those of low-mass SDSS galaxies.  While oxygen abundances of the 
LG dwarf galaxies are measured via spectroscopy of \HII regions 
(M98), abundances for SDSS galaxies were estimated
using empirically and theoretically calibrated relations between 
metallicity and strong optical emission line flux.
\citet{tre04} notes that aperture bias can lead to overestimates 
in abundances by about 0.1 dex.  
They also cite evidence from \citet{ken03} that 
``strong line'' methods such as theirs may overestimate the true 
abundance by as much as a factor of 2.  Taking these overestimates
into account, the LG dwarf galaxies coincide with the lower
end of the SDSS distribution and extend the mass range down by about two 
magnitudes.

\subsection{Star Formation Rate and Mass}
\label{sMsfr}

\begin{figure}
\epsscale{1.0}
\plotone{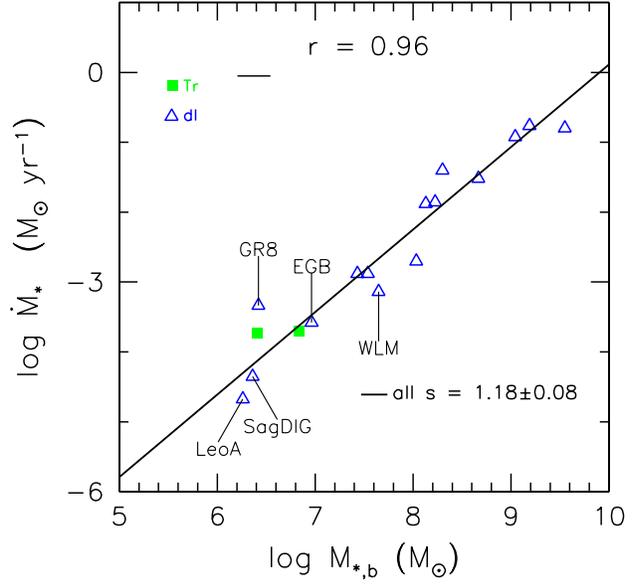}
\caption{Star formation rate vs. stellar mass.  
Shown is the two-dimensional linear regression 
with its slope $s$
indicated as well as the correlation coefficient $r$.}
\label{sfr}
\end{figure}

\Fig{sfr} shows the SFR versus $\Ms$ for dI's only (and two Tr's; dE's
have very little or no star formation).
The best-fitting scaling relation is $\sfr \prop \Ms^{\scsfr}$ with $r=\rsfr$.

\cite{bri04} (see their figure 17) 
showed that the low-mass SDSS galaxies follow a SFR-stellar mass relation
with a log slope of $\sim$0.7, noticeably shallower than the LG dwarf
galaxies.  Moreover, the LG dwarf galaxies fall significantly below
(almost one order of magnitude at least) the SDSS distribution.  
Adjusting for the IMF differences in the SFR determinations 
of M98 (Salpeter IMF) and \cite{bri04} (Kroupa IMF) would only
{\it increase} the vertical offset between them.
Since dwarf galaxies have very low metallicities, which normally means
very low dust content, the effects of dust extinction would also be
stronger at higher mass.  Correcting for this would further 
steepen the LG relation.  Underestimation of the Balmer absorption in
the H$\alpha$ maps which are used to estimate SFR is also more likely
at higher mass (Brinchmann 2005, private communication), and 
correcting for this would also lead to a steepening of the relation.
The different slope and vertical offset between the SDSS and LG
galaxies seems to be real, and present a new theoretical challenge.
We discuss this further in \sec{spediscuss}.

\subsection{PCA: The Fundamental Line of the Structural + SF
  parameters}
\label{strsepca}

\begin{figure*}
\epsscale{2.2}
\plottwo{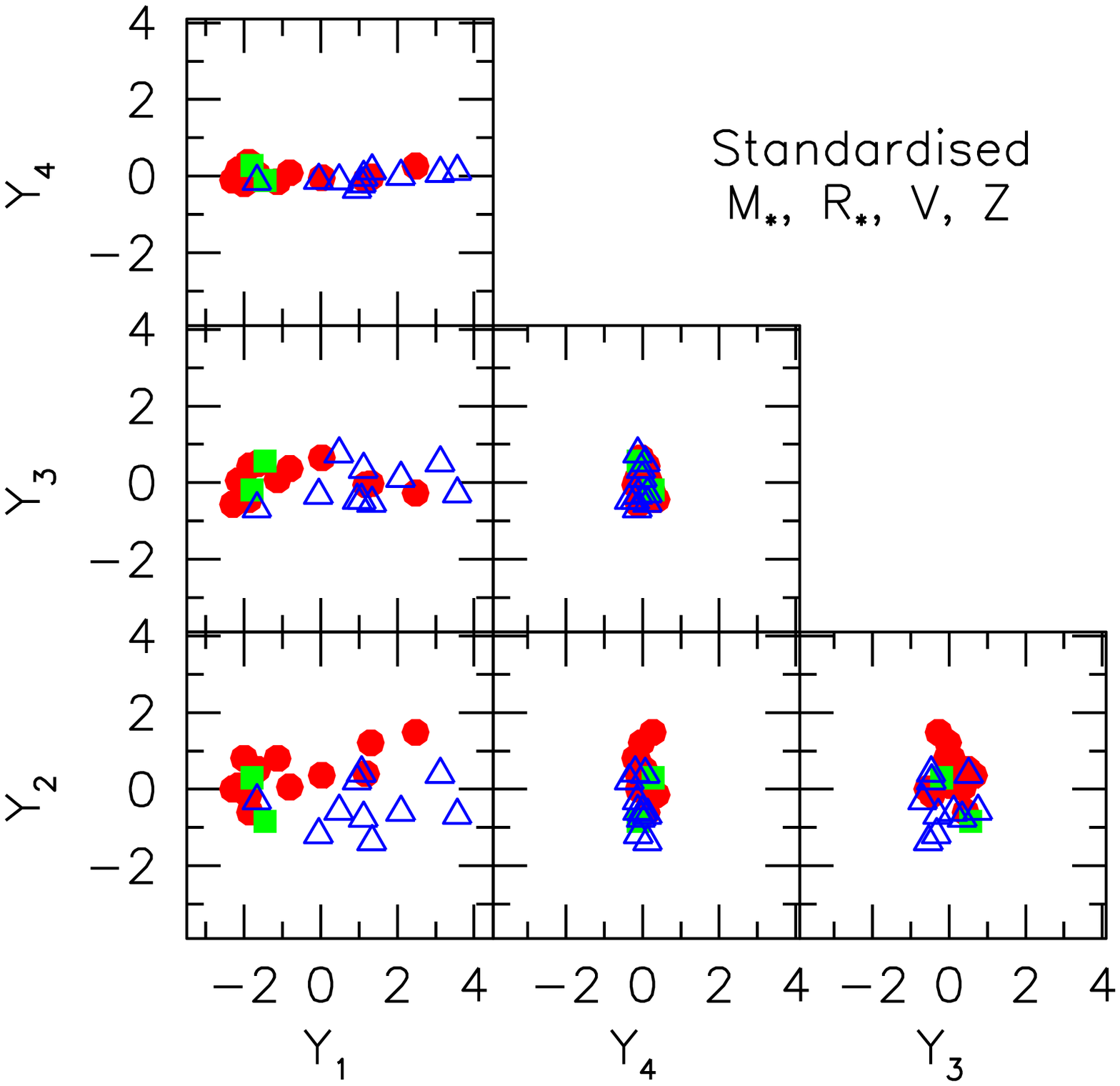}{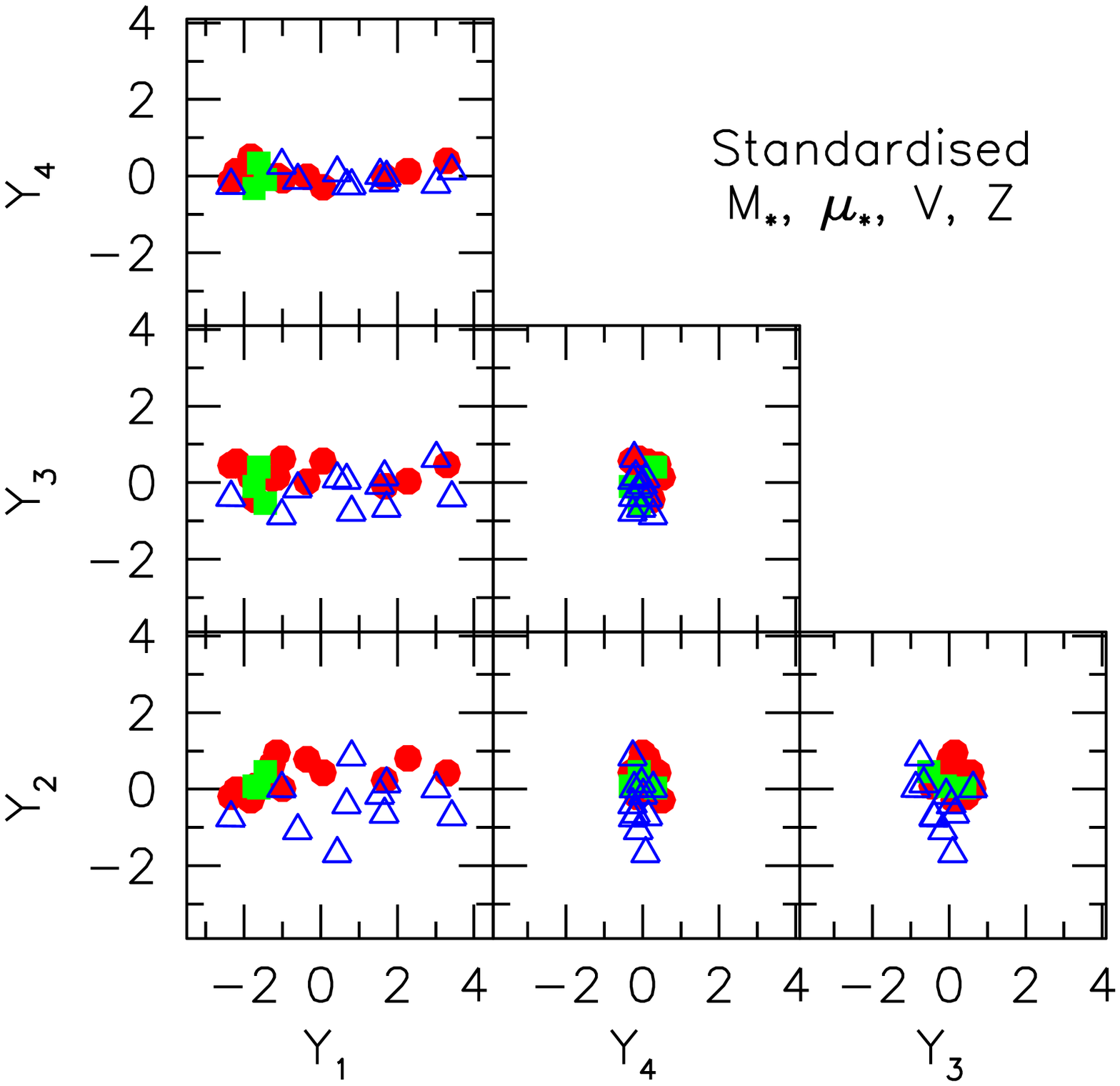}
\caption{The fundamental line in structural space + metallicity
  ($\Ms,\mus,V,Z$). 
The eigenvalues of the PCA are given in \tab{astrzpca}.
}
\label{strzfig}
\end{figure*}

The above linear correlations motivate an investigation of the
fundamental line in higher dimensional spaces than the structural
spaces that we have already analysed.  Thus, adding the metallicity 
to the structural spaces, we performed PCA on the 4-D (log) space 
of $\Ms,\Rs,V,Z$ and on $\Ms,\mus,V,Z$ space, and find that the 
fundamental line remains linear in these spaces in that the first 
eigenvalue is much larger than the other three.  The line is 
is illustrated in \fig{strzfig} 
and the eigenvalues are listed in \twotabs{strzpca}{astrzpca}.

However, a glance at
\fig{strzfig} reminds us that the dE's and dI's are offset in 
metallicity and suggests that they separately lie in linear
distributions.  Thus we also perform PCA on the dE's and dI's
separately.

\begin{figure}
\epsscale{1}
\plotone{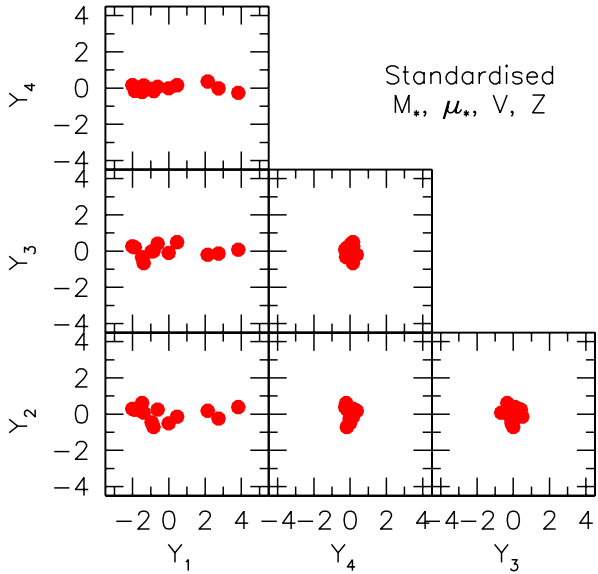}
\caption{Projections of the dE data in structural space + metallicity
  ($\Ms,\mus,V,Z$). 
The eigenvalues of the PCA are given in \tab{astrzpca}.
}
\label{strzfigs}
\end{figure}

For the dE's, we find that their fundamental line lies in the same
space as the fundamental line of the whole group but the dE
fundamental line is much tighter as indicated by their eigenvalues
(listed in \twotabs{strzpca}{astrzpca}),
especially in $\Ms,\mus,V,Z$ space (displayed in \fig{strzfigs}). 

From looking at the ratios of the eigenvalues, the dI's appear oblate
in $\Ms,\Rs,V,Z$ space, and
"quadraxial" in $\Ms,\mus,V,Z$ space.  However noting that the first eigenvalue
is much larger than the others in both these spaces, the distribution of the data 
extends mostly in one direction, and is to first order linear.  
Hence this distribution is also characterised
by one primary parameter.
Recalling the tight correlation between $\Ms$ and $\sfr$ in
\sec{sMsfr} for dI's, we add $\sfr$ to the parameter space
and
perform the PCA.  We find that the dI's are linearly distributed in
this parameter space (see \fig{sfrdi}), as indicated by their
eigenvalues in \tab{sfrdipca}.  

Thus we find that, when the stellar evolution parameters $Z$ and $\sfr$
are considered together with the structural parameters, the distribution
of all the LG dwarf galaxies remains linear.  In particular, the dE's 
and dI's separately form tighter linear distributions in the parameter 
space of $\Ms,\mus,V,Z$, and also of $\Ms,\mus,V,Z,\sfr$ for the dI's.

\begin{table}
\centering
\begin{minipage}{87mm}
\setlength{\tabcolsep}{0.9mm}
\caption{Eigenvalues for PCA in the parameter space of $\Ms$, $R$, $V$, $Z$\label{strzpca}}
\begin{scriptsize}
\begin{tabular}{ r l r l r l }
\hline\hline
\mcol{2}{c}{All galaxies} & \mcol{2}{c}{dE's only} & \mcol{2}{c}{dI's only} \\
\colh{Values (\%)} & \colh{Ratios} & \colh{Values (\%)} & \colh{Ratios} & \colh{Values (\%)} & \colh{Ratios} \\
\hline
 81.4 $\pm$  3.9  & \mrow{2}{*}{\}  6.2$\pm$ 1.6   } &  84.8 $\pm$  6.4  & \mrow{2}{*}{\}  7.6$\pm$ 3.2   } &  80.9 $\pm$  8.9  & \mrow{2}{*}{\}  7.3$\pm$ 4.0   } \\
 13.2 $\pm$  3.4  & \mrow{2}{*}{\}  2.8$\pm$ 1.1   } &  11.1 $\pm$  4.5  & \mrow{2}{*}{\}  3.3$\pm$ 2.9   } &  11.1 $\pm$  6.0  & \mrow{2}{*}{\}  1.4$\pm$ 1.0   } \\ 
  4.6 $\pm$  1.2  & \mrow{2}{*}{\}  6.4$\pm$ 2.5   } &   3.4 $\pm$  2.7  & \mrow{2}{*}{\}  4.7$\pm$ 4.4   } &   7.7 $\pm$  3.7  & \mrow{2}{*}{\} 23.4$\pm$15.8   } \\ 
  0.7 $\pm$  0.2  & &   0.7 $\pm$  0.4  & &   0.3 $\pm$  0.2  & \\
\hline\hline
\end{tabular}
\end{scriptsize}
\end{minipage}
\end{table}

\begin{table}
\centering
\begin{minipage}{87mm}
\setlength{\tabcolsep}{0.9mm}
\caption{Eigenvalues for PCA in the parameter space of $\Ms$, $\mus$, $V$, $Z$\label{astrzpca}}
\begin{scriptsize}
\begin{tabular}{ r l r l r l }
\hline\hline
\mcol{2}{c}{All galaxies} & \mcol{2}{c}{dE's only} & \mcol{2}{c}{dI's only} \\
\colh{Values (\%)} & \colh{Ratios} & \colh{Values (\%)} & \colh{Ratios} & \colh{Values (\%)} & \colh{Ratios} \\
\hline
 84.3 $\pm$  3.9  & \mrow{2}{*}{\}  8.6$\pm$ 2.9   } &  92.4 $\pm$  6.2  & \mrow{2}{*}{\} 22.4$\pm$23.3   } &  81.9 $\pm$  8.1  & \mrow{2}{*}{\}  5.7$\pm$ 3.1   } \\ 
  9.8 $\pm$  3.3  & \mrow{2}{*}{\}  2.1$\pm$ 0.9   } &   4.1 $\pm$  4.3  & \mrow{2}{*}{\}  1.6$\pm$ 2.0   } &  14.4 $\pm$  7.7  & \mrow{2}{*}{\}  4.5$\pm$ 3.2   } \\ 
  4.7 $\pm$  1.2  & \mrow{2}{*}{\}  3.8$\pm$ 1.5   } &   2.6 $\pm$  1.9  & \mrow{2}{*}{\}  3.0$\pm$ 2.6   } &   3.2 $\pm$  1.4  & \mrow{2}{*}{\}  6.4$\pm$ 4.0   } \\ 
  1.2 $\pm$  0.4  & &   0.9 $\pm$  0.4  & &   0.5 $\pm$  0.2  & \\
\hline
\end{tabular}
\end{scriptsize}
\end{minipage}
\end{table}

\begin{table}
\centering
\begin{minipage}{51mm}
\caption{Eigenvalues for PCA on in the parameter space of $\Ms$, $\mus$, $V$, $Z$, $\sfr$\label{sfrdipca}}
\begin{scriptsize}
\begin{tabular}{ r l }
\hline\hline
\mcol{2}{c}{dI's only} \\
\colh{Values (\%)} & \colh{Ratios} \\
\hline
 83.9 $\pm$  7.2  & \mrow{2}{*}{\}  6.9$\pm$ 3.7   } \\ 
 12.2 $\pm$  6.6  & \mrow{2}{*}{\}  4.7$\pm$ 3.4   } \\ 
  2.6 $\pm$  1.3  & \mrow{2}{*}{\}  2.5$\pm$ 1.6   } \\ 
  1.1 $\pm$  0.5  & \mrow{2}{*}{\}  4.1$\pm$ 2.5   } \\ 
  0.3 $\pm$  0.1  & \\ 
 \hline 
\end{tabular} 
\end{scriptsize} 
\end{minipage}
\end{table}

\section{DISCUSSION}
\label{conc}

We have studied the scaling relations of the dwarf galaxies in
physical quantities and quantified the ``fundamental line'' (FL) in
the parameter space of these quantities.  The following summarises 
how we achieved those goals and our results.

\subsection{The Data}
To derive the physical stellar mass, we used a combination of two
methods to derive stellar mass-to-light 
ratios.  These methods involved
the colour-$\MSL$ relations of \cite{bel01} (updated by
\citealp{bel03}), and SF histories of LG galaxies
coupled to population models of BC03.  

We divided the six parameters
($\Ms$, $\Rs$, $\mus$, $V$,
$Z$, $\sfr$) into two categories.  The first
category is the ``structural'' quantities ($\Ms$, $\mus$, $\Rs$, $V$), and
the second is the ``star formation''
(SF) quantities
($Z$, $\sfr$).  We treat the stellar mass as the independent
parameter and determine the scaling relations of the other
quantities with respect to it.  We list these scalings 
in \tab{sscalings}.   
Although we have used only the bisector fits, for
comparison with other studies, we
include in this table the slopes for the 1-D forward least
squares fits.

\begin{figure}
\epsscale{1}
\plotone{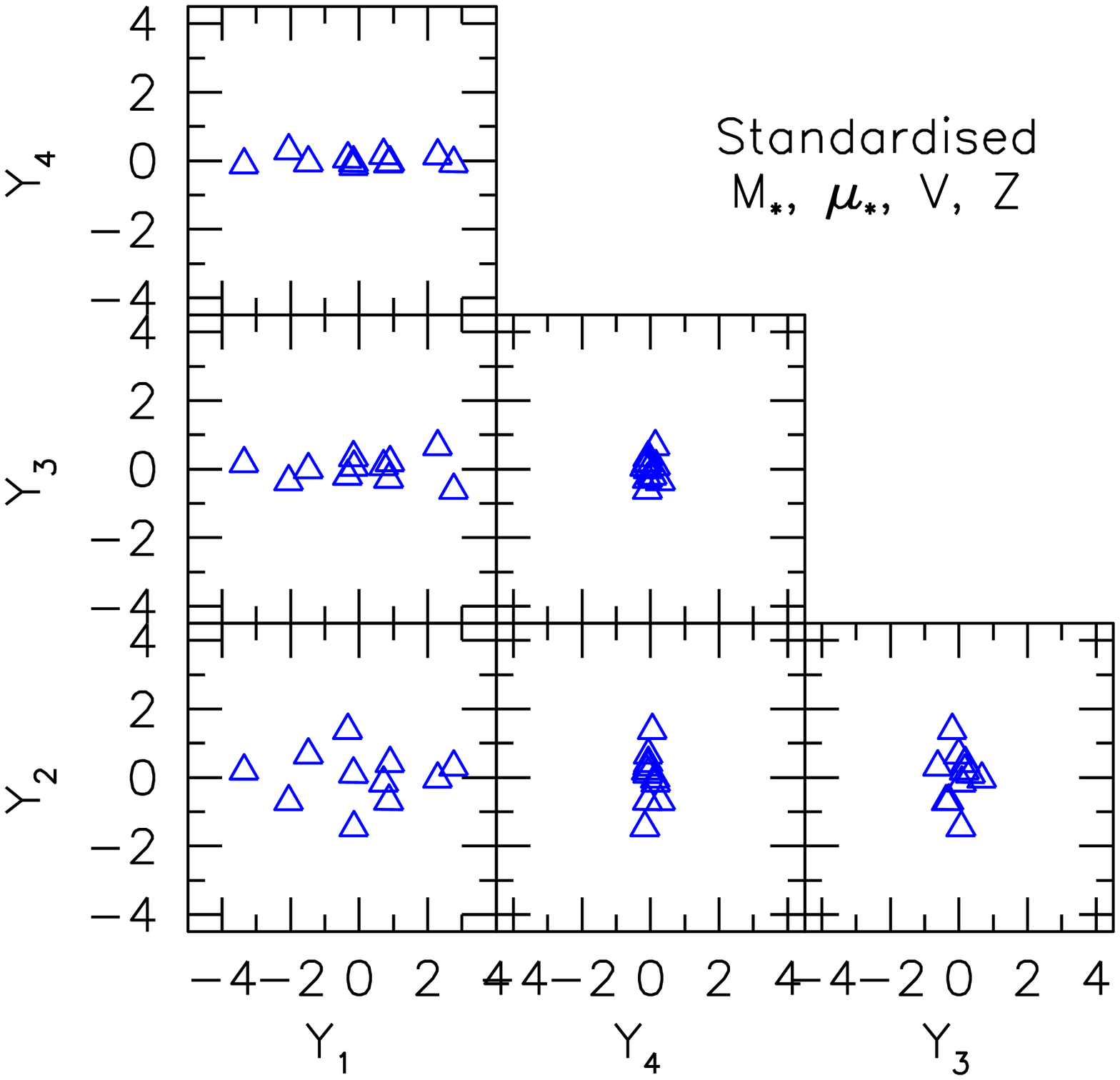}
\caption{Projections of the dI data in structural space + metallicity  
  ($\Ms,\mus,V,Z$). 
The eigenvalues of the PCA are given in \tab{astrzpca}.
}
\label{strzfigi}
\end{figure}
 
\begin{figure}
\epsscale{1}
\plotone{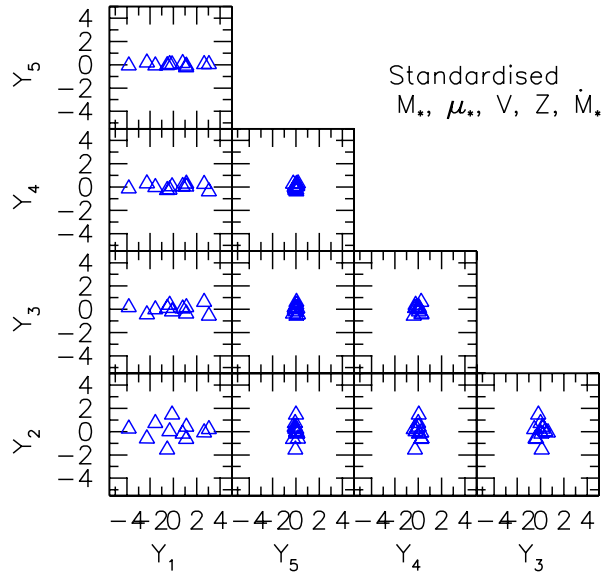}
\caption{Projections of the LG dI's in parameter (log) space of
  $\Ms$, $\mus$, $V$, $Z$ and $\sfr$.  
The components of the PCA eigenvectors are given in \Tab{sfrdipca}.
}
\label{sfrdi}
\end{figure}

\subsection{The Structural Scaling Relations}

We find that the scaling relations for LG dwarf galaxies
extend the corresponding relations found
in larger galaxies. 
Among the structural scaling relations, 
the $\Ms$-$\mus$ relation of the LG dwarfs match that
of the lower-mass regime in the SDSS, and extend the range of
the relation to about $10^6 \Msun$.  
This relation is also consistent with that found by
\cite{vad05b} after converting their photometric data to stellar
quantities.  The $\Ms$-$\Rs$
for the LG dI galaxies also seem to extend the corresponding
relation for large disc galaxies \citep{cou07}, and that of the
late-type galaxies in the SDSS \citep{she03} but with
steeper slope.  

The slope of the TF relation for the LG dwarf galaxies 
is consistent with that found by \cite{mcg05}
though we do not find their reported
break at 90 $\kms$.  The LG dwarf galaxies follow a smooth TF relation with no
indication of a mass or velocity scale.  When adding the \HI mass, 
the LG dI's follow a BTF relation that matches that
of \cite{mcg05} above $V=32 \kms$, with reduced scatter.  However,
including all the dI's yields a shallower slope compared to the
optimal slope of \cite{mcg05}, but consistent with BdJ and 
\cite{geh06}.  \cite{mcg05} contended that the BTF slope is sensitive
to the method used to determine $\MSL$, so our shallower slope is expected.

\subsection{The SF Scaling Relations}
\label{spediscuss}

The $\Ms$-$Z$ relation also appears
to be a low-mass extension of the relation for the SDSS galaxies.  

The slope of the $\sfr$-$\Ms$ relation for the LG dI's ($\scsfr$) is 
significantly steeper than that of the SDSS star-forming galaxies
\citep[0.7,][]{bri04}.  Additionally, the LG dI's fall at least one order of
magnitude below the SDSS distribution.  Correcting for IMF differences
and observational effects such as dust and Balmer absorption
worsens this discrepancy.  This real vertical offset and steeper
slope suggests
some environmental effect on current star formation.  

\begin{figure}
\epsscale{1.0}
\plotone{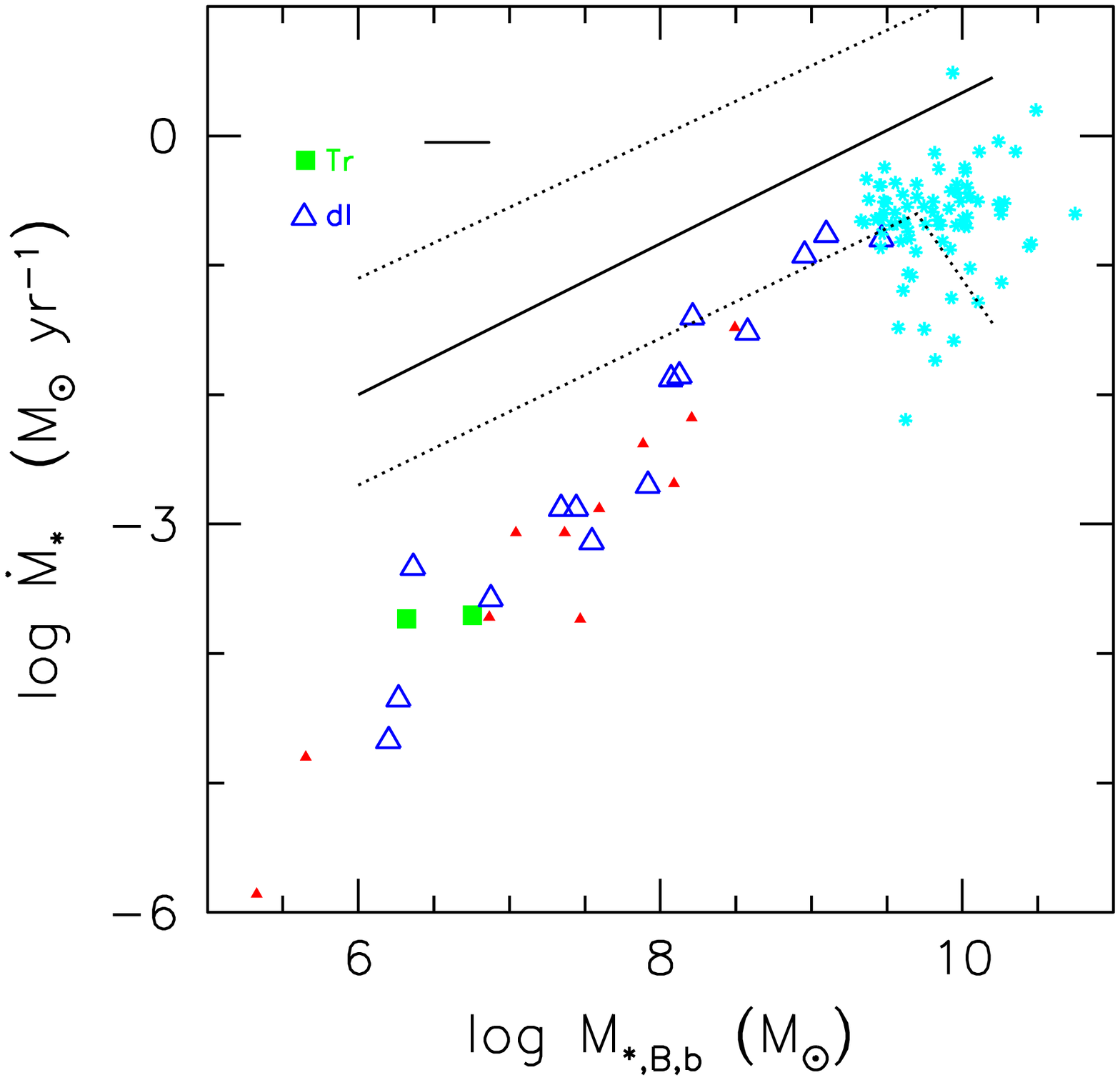}
\caption{SFR data for the Sculptor Group dwarf galaxies (solid red
  triangles) 
  and the
  Coma Cluster dwarf galaxies (cyan stars) plotted together with the LG dI
  and Tr galaxies (blue and green).  Data for the Sculptor Group is
  taken from \citet{ski03}, and for the Coma Cluster, we used 
  \citet{gav98}.  We assume a $\MSL_B$ value of 0.6, and adopt the 
  prescription of \citet{ken98} to estimate SFR from H$\alpha$ measurements. 
  The solid line is roughly (estimated by eye) the mode line of the distribution of SFR in bins of log $\Ms$ for 
  the SDSS, while the dotted lines roughly mark the lines where the conditional likelihood of SFR 
  a given $\Ms$ is equal to 0.02 \citep[see Fig. 17 of][]{bri04}.
}
\label{compSFR}
\end{figure}

Using the
SFR data from \cite{ski03} for the Sculptor Group dwarf galaxies, and a
$\MSL_B$ value of 0.6 (the average of the $B$-band SFR and BdJ
values from \tab{avgML} for dI's) to estimate their stellar
masses, we find that the Sculptor Group dwarfs lie along the
same $\Ms$-SFR relation as LG dwarfs.  In addition, using
the H$\alpha$ data from \cite{gav98} for the Coma cluster dwarf
galaxies, the \cite{ken98} relation between H$\alpha$ and SFR , and the same
$\MSL_B$ value to estimate their $\Ms$, we find that the Coma
cluster dwarf galaxies roughly lie on the SFR-$\Ms$ relation of the
SDSS (see \fig{compSFR}). 
Since we expect that the average
SDSS galaxy lies in denser environments than the Local and Sculptor Groups,
these observations suggest that the SFR-$\Ms$
relation for dwarf galaxies depends on environment.

\subsection{The ``Fundamental Line'' of the LG Dwarf Galaxies}

We have used PCA to quantify the distribution of the LG dwarf galaxies
in the parameter (log) space of the structural quantities alone, and
also of the structural + SF quantities.  We make use of
the output of PCA to display the full dimensionality and extent of
the distribution of the galaxies from the most useful orthogonal angles.  

Using PCA, we find that the LG dwarf galaxies (all types) form one linear
distribution in structural space $\Ms,\mus$ (or $\Rs$), $V$.  Despite
the fact that dE's tend to have higher $Z$ than dI's, 
the distribution of the entire dwarf population
remains linear when we add $Z$ to the parameter space.
Since this fundamental line resides in the same space with consistent slopes
as those predicted by supernova feedback models 
($\Ms,\mus$ (or $\Rs$), $V,Z$ space), we 
call this space ``SNF'' space \citep{dek03}.

Since the star formation properties of dE's and dI's differ appreciably,
especially $Z$, we performed PCA on dE's and dI's
separately.  We find that the dE's form an even tighter
linear distribution in SNF space.   The dI's are also linear in SNF space, 
and are even tighter in the space spanned by
$\Ms,\mus,V,Z,\sfr$.  

The linearity of these distributions suggest that
one primary parameter governs their physics.  Furthermore because dI's tend
to be more massive than the dE's, halo mass may be one of the 
drivers of the distinction between the two types. 

\bigskip
{\it Comparison with Previous Work}
\medskip

PB02 found a ``fundamental line'' in the parameter space (to which we will
refer as ``PB'')
of total mass-to-light ratio $\MtLv$, metallicity [Fe/H], and central
surface brightness $\Sigma_V$ \citep[see also][]{sim06}.  
For $\MtLv$, PB02 used total masses, estimated
dynamically, and total luminosity.  PB space can be
derived from SNF space through the following relations:
\begin{eqnarray}
\MtLv &=& V^2 R /L \prop V^2 R /\Ms \prop V^2 / \sqrt{\Ms \mus}\nonumber \\
\Sigma_V &\prop& -2.5 \log \mus \nonumber \\
{\rm [Fe/H]} &\prop& \log Z.
\label{ptransf}
\end{eqnarray}

PB space is simply SNF space collapsed into three dimensions.  We have 
improved on their analysis in two ways.  Firstly, we included more dI's  
in our sample which contribute enough scatter   
so that it is no longer linear in PB space.  Secondly, we have broken PB 
space into three independent structural parameters ($M$, $V$, and $R$),
and metallicity $Z$.  PB space relates the three 
structural parameters by only one relation, namely 
$\log \MtLv \prop 0.67 \log \mus$ (PB02), which translates to one
projection of the distribution.  A linear distribution
in the three-space of the structural parameters should be described by
{\it two} orthogonal projections which we graphically and
quantitatively present.  

\begin{table*}
\centering
\begin{minipage}{108mm}
\caption{\small Scaling Relations\label{sscalings}}
\begin{scriptsize}
\begin{tabular}{l c c c l c c c}
\hline\hline
\colh{vs. $\logMstarv_b$}&\colh{$\sbis$}& \colh{$\bbis$} & \colh{$\sigbis$} & & \colh{$\s1d$}& \colh{$\b1d$} & \colh{$\sig1d$} \\
\hline
\mcol{8}{l}{\bf All LG dwarf galaxies}\\
$\mustarv_b$ &   0.61 $\pm$  0.05 &   2.94 $\pm$  0.35 &   0.38 & &   0.49 $\pm$  0.05 &   3.80 $\pm$  0.35 &   0.41 \\                                
$\logrp$ &   0.33 $\pm$  0.02 &  -2.81 $\pm$  0.15 &   0.21 & &   0.28 $\pm$  0.03 &  -2.44 $\pm$  0.20 &   0.21 \\                                    
$\logv4$ &   0.27 $\pm$  0.01 &  -0.51 $\pm$  0.11 &   0.10 & &   0.26 $\pm$  0.02 &  -0.39 $\pm$  0.12 &   0.11 \\                                    
$\logZ$ &   0.40 $\pm$  0.04 &  -6.07 $\pm$  0.30 &   0.27 & &   0.29 $\pm$  0.04 &  -5.33 $\pm$  0.26 &   0.28 \\                                     
\mcol{8}{c}{ }\\
\mcol{8}{l}{\bf dE galaxies}\\
$\mustarv_b$ &   0.74 $\pm$  0.05 &   2.25 $\pm$  0.37 &   0.27 & &   0.66 $\pm$  0.06 &   2.76 $\pm$  0.40 &   0.35 \\                                
$\logrp$ &   0.26 $\pm$  0.05 &  -2.36 $\pm$  0.30 &   0.15 & &   0.17 $\pm$  0.03 &  -1.80 $\pm$  0.20 &   0.20 \\                                    
$\logv4$ &   0.29 $\pm$  0.02 &  -0.59 $\pm$  0.17 &   0.09 & &   0.27 $\pm$  0.03 &  -0.47 $\pm$  0.20 &   0.15 \\                                    
$\logZ$ &   0.40 $\pm$  0.05 &  -5.87 $\pm$  0.29 &   0.10 & &   0.37 $\pm$  0.05 &  -5.68 $\pm$  0.33 &   0.20 \\                                    
\mcol{8}{c}{ }\\
\mcol{8}{l}{\bf dI galaxies}\\
$\mustarv_b$ &   0.70 $\pm$  0.08 &   2.01 $\pm$  0.69 &   0.37 & &   0.51 $\pm$  0.09 &   3.52 $\pm$  0.81 &   0.53 \\                                
$\logrp$ &   0.41 $\pm$  0.04 &  -3.44 $\pm$  0.41 &   0.20 & &   0.32 $\pm$  0.04 &  -2.69 $\pm$  0.39 &   0.27 \\                                    
$\logv4$ &   0.26 $\pm$  0.02 &  -0.43 $\pm$  0.16 &   0.08 & &   0.24 $\pm$  0.02 &  -0.29 $\pm$  0.16 &   0.14 \\                                    
$\logZ$ &   0.49 $\pm$  0.06 &  -7.09 $\pm$  0.49 &   0.24 & &   0.43 $\pm$  0.05 &  -6.64 $\pm$  0.42 &   0.27 \\                                    
$\logsfr$ &   1.18 $\pm$  0.08 & -11.69 $\pm$  0.61 &   0.25 & &   1.14 $\pm$  0.08 & -11.35 $\pm$  0.66 &   0.35 \\                                   
\hline
\end{tabular}
\end{scriptsize}
\end{minipage}
\end{table*}

\cite{ber03} studied the fundamental plane of SDSS early-type galaxies.
They find that these galaxies obey
$R \prop \sigma^{\sim 1.5} I^{\sim 0.75}$, with slight
variations between filters, where $I=(L/2 R^{-2})$ and $\sigma$ is
the velocity dispersion.  Although the plane equation describing the
LG dE's $\Rs \prop V^{2.1 \pm 0.7} \mus^{0.6 \pm 0.2}$ is consistent
with that of the SDSS early-type galaxies, 
the LG dE's are more accurately a linear
distribution in the space spanned by $\Rs,\sigma,\mus$ (a permutation
of $\Ms, V, \mus$ structural space).  
This is consistent with the findings of
\cite{ben92} who show the line of dE's protruding from the plane of
elliptical galaxies in ``$\kappa$''-space, which is
a permutation of structural space.

\cite{zar06b} \citep[and later ][]{zar07} find that the LG dE's lie along a ``fundamental
manifold'' of spheroidal galaxies in the space spanned by $R_e$,
$\sigma$, $I_e$ and $M/L$.  (They note that log$M/L \prop \log^2 \sigma$ 
and so add the non-power-law term $\log^2 \sigma$ to
their plane equation instead of log$M/L$.)
One can easily reduce these four parameters to 
three independent structural parameters ($L$, $\sigma$ and $R$ or
$\mu$) through \eq{ptransf} spanning a 3D space in which the LG
dE's form an even tighter line.  

So in summary, while the dI's seem to extend the structural scaling
relations of giant late-type galaxies to lower masses, the
dE's seem to depart from the fundamental plane of giant early-type
galaxies to form a distribution described by a one-parameter family of
equations.

\section{CONCLUSION}

Our analysis of structural and spectroscopic data for LG galaxies 
has provided important constraints for modelling the physical processes
that govern the formation of dwarf galaxies.  In particular,
successful models will need to reproduce the fundamental line in
structural space +$Z$ (\ie SNF space) obeyed by all dwarf galaxies.
One such model is the simple supernova feedback prescription of 
\cite{dek03} which quite accurately predicts the slopes of the scaling
relations using $\Ms$ as the one primary parameter determining the 
distribution of the data.

What supernova feedback does not deal with is the zero-points of the
scaling relations, particularly the different zero-points between
early and late-type dwarf galaxies.
Future successful models must include secondary physical processes
that distinguish between early and late-type dwarf galaxies, producing
the separate higher-dimensional fundamental lines that they are
observed to follow in SNF and $\Ms,\mus,V,Z,\sfr$ spaces.
The graphical display of PCA output as we have 
done in this paper can also help in visualising and  
understanding scatter in model dwarf galaxies.

Several unexplained challenges 
include: 
\begin{itemize}
\item
the different normalisations of the metallicity and surface density scaling
relations for different Hubble type;
\item
the presence of recent star formation episodes in late type galaxies, 
and hence the secondary fundamental line that dI's
follow (\ie in the parameter space of $\Ms,\mus,V,Z,\sfr$); and 
\item
the dependence of the dwarf galaxy SFR scaling relation with environment,
particularly the steepening of the relation in lower density
environments, and the order of magnitude drop in SFR in lower density
environments.
\end{itemize}

\section*{Acknowledgements}
We acknowledge helpful and stimulating discussions with
Jarle Brinchmann, Marla Geha, Eva Grebel, Guinevere Kauffmann, 
and Lauren MacArthur, as well as Ofer Lahav for sharing an early 
version of his PCA code.  We also thank Dennis Zaritsky for pointing
out an error in a draft version.  This research has been supported by 
the National Science and Engineering Council of Canada, by 
the Israel Science Foundation grant 213/02 
by the German-Israel Science Foundation grant I-895-207.7/2005, 
by the France-Israel Teamwork in Sciences 2008,
by the Einstein Center at HU,
and by a NASA ATP grant at UCSC.

\bibliographystyle{mn2e}
\bibliography{jobib}

  \renewcommand{\theequation}{A-\arabic{equation}}
  \renewcommand{\thesection}{A-\arabic{section}}
  \setcounter{equation}{0}  
  \setcounter{section}{0}  
  \section*{APPENDIX A: $\MSL$ from SSP Models}  
\label{app:bc}

\citet[hereafter M98]{mat98} has attempted to reconstruct the relative 
star formation rate (SFR) as a function of age for many of the dwarf 
galaxies.  We define a function $\Psi(t)$ which is simply the SFR 
as a function of $t$, where $t=0$ is the time of the first burst
of star formation, and call $\Psi(t)$ the ``star formation history'' 
or SFH.  We normalize $\Psi(t)$ such that its integral over all time is 1.

\citet[hereafter BC03]{bru03} have modelled the spectral
evolution of simple stellar populations (SSP), \ie instantaneous bursts of 
star formation, using evolutionary tracks from their 
``Padova 1994 library'' and 
\citet{cha03} IMF (which produces very similar results for $\MSL$ as
the Kroupa IMF - see their paper).  BC03 give stellar mass $\Ms(t)$ and 
absolute $V$ magnitudes as a function of time of a stellar population 
with a mass normalised to 1$\Msun$ at the time of the burst, 
for a population of a given metallicity $Z$.  
Model results cover the range 
$Z$=0.0001, 0.0004, 0.004, 0.008, 0.02, and 0.05.  

However, $Z$ is also a function of time.  \citet{bin98}
describe different chemical evolution models where $\dot{Z}$ is roughly
proportional to $\dot{\Ms}/M_{gas}$.  To simplify the calculation, 
we adopt $\dot{Z} \prop \dot{\Ms}$, with 
\be
Z(t) = Z_{\rm tot} \int_{t_{\rm o}}^t \Psi(t'){\rm d}t'
\label{amet}
\ee
where $Z_{\rm tot}$ is from the measured metallicity (\eq{metal} in
the main text).
Since $M_{gas}$ declines with time as the gas becomes stars, 
$Z(t)$ increases somewhat faster
with the time of the burst than our simplified $Z(t)$ in \eq{amet}.
However, our resulting $\MSL$ ratios are not very different from the model
with constant $Z$ (the largest difference in log $\MSL$ being 0.05, or 
less than 1\% of log $\Ms$) and we conclude that our simplified
function $Z(t)$ will not significantly overestimate $\Ms$. 

Given a SFH of a galaxy, $\Psi(t)$, we can use the BC03 SSP 
models to predict the final 
total luminosity $L(t_{\rm o})$ of the galaxy from a convolution
of the SFH with the luminosity due to $V(t,Z(t))$:
\be
L_{\lambda} = \int_0^{t_{\rm o}} \Psi(t')l_{\lambda}(t',Z(t'))\;{\rm d}t'
\ee
where $t=t_{\rm o}$ is the present,
and $l_{\lambda}(t,Z(t))$ is the 
luminosity of the burst for a particular band $\lambda$
in physical solar units such that: 
\be
l_V(t,Z(t)) = 10^{0.4[{\rm M}_{V,\odot}-V(t,Z(t))]}.
\ee
$\Psi(t)$ is normalised
\be
\int_0^{t_{\rm o}} \Psi(t)\;{\rm d}t = 1
\ee
so that $L_{\lambda}$ is the total light in solar units for the star burst.

Similarly, the total stellar mass is:
\be
M_{*,tot} = \int_0^{t_{\rm o}} \Psi(t')\Ms(t_{\rm o}-t',Z(t_{\rm o}-t'))\;{\rm d}t'.
\ee


  \renewcommand{\theequation}{B-\arabic{equation}}
  \renewcommand{\thesection}{B-\arabic{section}}
  \setcounter{equation}{0}  
  \setcounter{section}{0}  
  \section*{APPENDIX B: The Mechanics of PCA}
\label{app:mech}

Given a data set with $m$ parameters, 
the PCA quantifies the distribution of the data 
by performing a series of rotations on these original basis vectors of
the parameter space.  

Before applying the PCA, the data are reduced in the following way:
If the original data set contains
$n$ galaxies and $m$ physical parameters, 
then we can construct an $n \times m$ data matrix:
\be
A_{i,j} \equiv x_{i,j};~~ i=1..n,~ j=1..m
\ee
To simplify our calculations, the matrix is centred about 
the means of each parameter:
\be
A^{c}_{i,j} \equiv x_{i,j}-\bar{x}_j.
\ee
In order to better estimate the strength of the correlations, the matrix
can be ``standardised'' by dividing the elements by the standard deviations
of the parameters:
\be
A^{c,s}_{i,j} = \frac{x_{i,j}-\bar{x}_j}{\sigma_j}
\label{stand}
\ee
where
\be
\sigma_j = \sqrt{\frac{\sum ^{n}_{i=1} (x_{i,j}-\bar{x}_j)^2}{n-1}}.
\ee
Then the covariance or correlation matrix is calculated:
\be
{\rm Cov}_{j,k}=\frac{1}{n-1} \sum^n_{i=1} A^{c}_{i,j} A^{c}_{i,k} 
\label{unweightedcov}
\ee
and
\be
{\rm Cor}_{j,k}=\frac{1}{n-1} \sum^n_{i=1} A^{c,s}_{i,j} A^{c,s}_{i,k}; \; j,k=1..m
\label{unweightedcor}
\ee
These are simply the Pearson correlation coefficients for the $m$ parameters.

After this reduction of the data, the covariance or correlation matrices are 
passed to the PCA.  
Let the matrix {\bf C} be either {\bf Cov} or {\bf Cor}.  Then PCA runs 
on {\bf C} so 
as to calculate the $m \times m$ matrices {\bf V} and {\bf D} such that
\be
{\bf C V} = {\bf V D}
\ee
where {\bf V} is the product of the Jacobi rotation matrices that give
the diagonal matrix {\bf D}.  The elements of the diagonal matrix $D_{j,j}$
are thus the eigenvalues of {\bf C}, and the column vectors {\bf V}$_j$ are 
the eigenvectors of {\bf C}.  The Jacobi rotations are performed by the 
subroutine {\tt jacobi} from 
{\it Numerical Recipes}, \citealp{pre92}, \S 11.1).  
Our program for PCA is 
a modified version of a routine kindly provided to us by Ofer Lahav 
\citep[described in][]{mad02}.
Our modifications allow for 
error estimates in the eigenvalues and vectors.

\subsection{Error Estimates in PCA}

The error estimates in PCA were estimated using a
``bootstrap'' 
method.  
If the original data set has $n$ galaxies, 
the ``bootstrap'' method creates $N_{\rm B}$ new sets of data 
$\tilde{\bf A}_k$, $k=1,N_{\rm B}$
by randomly selecting $n$ galaxies from the original data set 
(allowing each galaxy a chance to be selected more than once).  Then the 
procedure described above is performed on the new data sets $\tilde{\bf A}_k$ 
to produce $\tilde{\bf V}_k$ and $\tilde{\bf D}_k$, 
$k = 1,N_B$. 
The bootstrap standard error is
\be
\sigma_{B,{\bf V}}^2 = \frac{1}{N_{\rm B}-1} \sum_{k=1}^{N_{\rm B}} (\tilde{\bf V}_k - \bar{\bf V}_B)^2
\ee
where $\bar{\bf V}$ is
\be
\bar{\bf V}_B = \frac{1}{N_{\rm B}} \sum_{k=1}^{N_{\rm B}} {\bf V}_k
\ee
Analogous relations yield 
$\sigma_{B,{\bf D}}$.

For our Bootstrap analysis, we used $N_B$ = 1000 to provide sufficient
statistics for estimating the bootstrap standard error.

\subsection{Projecting the Data in the New Basis}

The elements of the eigenvectors give the strength of the dependence of the 
vector on the original basis parameters.  The original data {\bf C}$_i$ for 
the $i$th galaxy (row) may be transformed by
\be
{\bf Y} = {\bf C}_i{\bf V}_j
\label{transf}
\ee
where the components of {\bf Y} can be seen as the coordinates in the new 
orthogonal base.  Thus for the unstandardised PCA, the components of the 
principal eigenvector give the scaling relations between the original 
parameters.  Since these scaling relations depend on the range of the 
original parameters, it is often useful to standardise the data according
to \Eq{stand} in order 
to eliminate any bias toward parameters with the largest range.

\label{lastpage}
\end{document}